\documentclass[aps,pre,twocolumn,groupedaddress]{revtex4-2}
\usepackage{amsmath,amssymb,graphicx,tikz,subfigure}
\usepackage[english]{babel}
\usepackage[utf8]{inputenc}
\usepackage{stmaryrd}

\newcommand{\sign}[0]{\text{sign}}

\begin{document}

\title{Quantum work statistics in regular and classical-chaotic dynamical billiard systems}

\author{Sebastian Rosmej}
\author{Mattes Heerwagen}

\affiliation{Carl von Ossietzky Universität Oldenburg, Institut für Physik, D-26111 Oldenburg, Germany}

\begin{abstract}
In the thermodynamics of nanoscopic systems the relation between classical and quantum mechanical description is of particular importance. To scrutinize this correspondence we have picked out two 2--dim billiard systems. Both systems are studied in the classical and the quantum mechanical setting. 
The classical conditional probability density $p(E,L|E_0,L_0)$ as well as the quantum mechanical transition probability $P(n,l|n_0,l_0)$ are calculated which build the basis for statistical analysis. 
We calculate the work distribution for a particle. Especially the results in the quantum case are of special interest since already a suitable definition of mechanical work in small quantum systems is controversial.
Furthermore we analysed the probability of both zero work and zero angular momentum difference. Using connections to an exact solvable system analytical formulas are given in both systems. In the quantum case we get numerical results with some interesting relations to the classical case. 

\end{abstract}

\maketitle

\section{Introduction}

One crucial ingredient for the thermodynamic characterization of small systems with typical energy turnover of the order of the thermal energy per degree of freedom is the statistical distribution of work \cite{Jar11,Sei12}. In the classical setting the definition of work is unambiguous. The work is defined as the integral of force along the trajectory. In the quantum case the definition of work meets some difficulties \cite{Yuk00,TalHae16,EngNol07,Binetal}.

There exists no analogue of the classical trajectory. A possible and intuitive way to define a quantum work is to measure the energy twice, before and after the process. This definition is called two projective measurement method \cite{Yuk00,TalHae16}. On the one hand, this definition is simple and operative. On the other hand, the measurements are likely to destroy quantum interferences that may be decisive for the non-classical behaviour of the system. To clarify which correlations are destroyed by the two projective measurement prescription and which are kept, it is instructive to look in detail at the correspondence between classical and quantum work distributions \cite{GarRonWis17a,GarRonWis17b}. This has been done for a quartic oscillator with time-dependent stiffness constant \cite{JarQuaRah15,Engetal20} and for a periodically driven quartic oscillator \cite{HeeEng20}. The latter system shows chaotic behaviour in some regions of parameter space. In this paper we extend the calculations to billiards with moving walls that are known to implement fully chaotic motion.

Billiards are common systems to study chaotic dynamics and have been investigated thoroughly since the pioneering work of Sinai \cite{Sin70} and Bunimovich \cite{Bun79} thoroughly. In two dimensions both regular and chaotic motions are possible. If in addition to the energy a conserved quantity exists, the dynamics is integrable. This is the case, e.g. for rectangles, circles and ellipses. Stadium billiards, on the other hand, are known to be chaotic.
In quantum mechanics the study of billiards with static walls became a central pillar in the theory of quantum chaos \cite{Eck88,Gut90,Sto00}. Situations with moving boundaries have been investigated much less. However, in recent years the interest in the classical dynamics of time-dependent billiards grows. Integrable time-dependent billiards were mainly discussed as toy-models for Fermi-acceleration \cite{LosRyaAki00,Lenetal09,Naketal11}. The results are very interesting, however, they all concern averages over a large number of particles. Quite generally, investigations of dynamical billiards that focus on the whole statistical distribution of energy are scarce \cite{GarRonWis17a,GarRonWis17b,Babetal15}. 

The aim of this paper is to compare the classical and quantum work statistics of dynamical billiards. Starting with one particle in a two dimensional circular billiard, we consider two systems based on two types of expansion: at first a dynamical billiard system due to an expanding radius (\textit{System 1}) and as a second a horizontal movement of the half circles in opposed directions building a Bunimovich stadium (\textit{System 2}) \cite{Bun79}. This expansion step will be followed by a contraction step up to the initial circle.

We prepare a system in equilibrium with a heat bath at inverse temperature $\beta = (k_\mathrm{B}T)^{-1}$. At time $t=0$ bath and system were decoupled. During the expansion and the contraction the system is isolated from the environment. Because of the first law of thermodynamics the work is given by the difference between initial energy $E_0$ and final energy $E$. The work statistics 

\begin{align}
\nonumber p&(W) = \\
&\int\limits_{0}^{\infty} dE \int\limits_{0}^{\infty} dE_0 ~p_0(E_0)p(E|E_0) \delta\left(W-\left(E-E_0\right)\right)
\end{align} 
involves the probability density $p_0(E_0)$ to start in the initial energy $E_0$ given by the Boltzmann distribution and the transition probability density $p(E|E_0)$ to end with energy $E$. $\delta(x)$ is the delta function.

The paper is organized as follows. In section \ref{sec:systems} we define \textit{System 1} and \textit{System 2} and describe them classically and quantum mechanically. Our results on the work statistics are given in section \ref{sec:results}. Therein we start with the transition probability, Sec. \ref{sec:trans}, which leads to the work distribution, Sec. \ref{sec:PW}. In Sec. \ref{sec:PW0} and \ref{sec:PL0} we focus on the probability of no energy change ($W=0$) and no angular momentum change ($\Delta L=0$). Finally, section \ref{sec:conc} contains our conclusion.

\section{Two dynamical Billiard systems} \label{sec:systems}

\begin{figure}
\centering
\includegraphics[width=\linewidth]{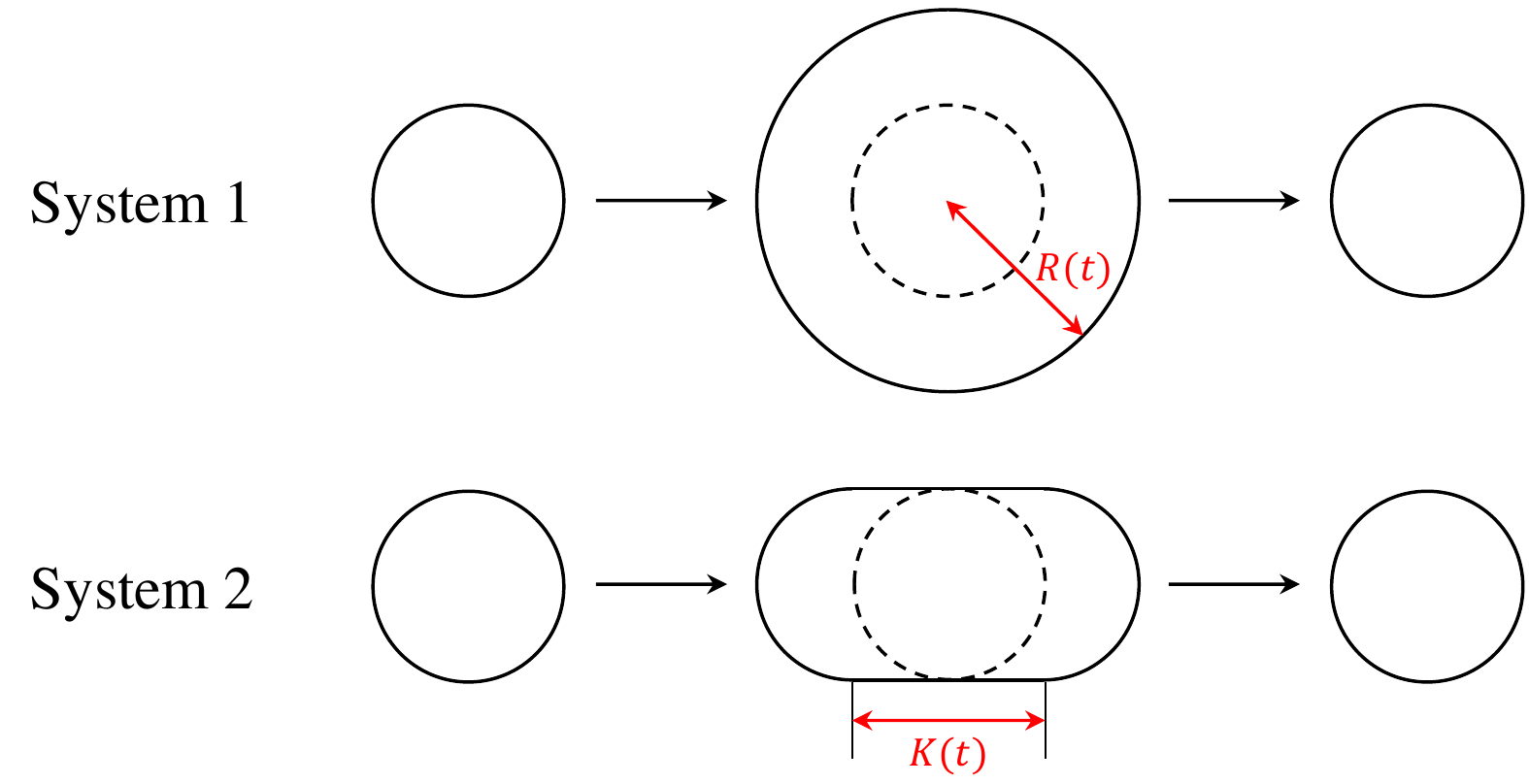}
\caption{The radially breathing circle (\textit{System 1}) and the horizontally breathing stadium (\textit{System 2}).}
\label{fig:systems}
\end{figure}

We start with a billiard system containing a two dimensional circular edge with radius $R_0$. At time $t=0$ the system starts to expand with a constant velocity $u>0$ up to time $T/2$ (expansion phase) followed by a contraction phase with a constant velocity $-u$ for $T/2 < t \leq T$ symmetric to the previous expansion. We consider two examples for the expansion and contraction phases: 
\begin{enumerate}
 \item[(i)] with a varying radius linearly with time 
  \begin{align}
	 \label{eq:Rt} R(t)=\begin{cases}R_0+u t & t \leq T/2 \\ R_0 + u(T-t) & T/2 < t \leq T \, , \end{cases}
	\end{align}
  in the following called \textit{System 1} and
 \item[(ii)] with a linear horizontal movement of the half circles in opposed directions forming a Bunimovich stadium billiard with an edge length 
  \begin{align}
	K(t)=\begin{cases} 2ut & t \leq T/2 \\ 2u( T-t) & T/2 < t \leq T \, , \end{cases}  
	\end{align}
	in the following called \textit{System 2}. 
\end{enumerate}
Both systems are illustrated in Fig.~\ref{fig:systems}. \textit{System 1} is a classically integrable system, whereas \textit{System 2} is chaotic. At $t=T$ both billiard systems end in the initial circular billiard with radius $R_0$. 

In the next subsections we first analyze the dynamics of classical particles in both systems and then the time-evolution of wave functions.




\subsection{Classical description} \label{sec:class}

For both systems we consider a classical particle starting at $t=0$ on position ${\bf r}_0=\left(\begin{smallmatrix} x_0 \\ y_0 \end{smallmatrix}\right)$ with the velocity ${\bf v}_0=\left(\begin{smallmatrix} a_0 \\ b_0 \end{smallmatrix}\right)$. The trajectory is a combination of rectilinear motions up to collisions with the edges. At each collision the particle velocity changes in modulus and direction. For both systems the collision times, positions and velocities are calculated analytically. In App.~\ref{sec:AppA} explicit formulas are given for \textit{System 1}, a short discussion is given for \textit{System 2}.

\begin{figure}
\centering
\includegraphics[width=.48\linewidth]{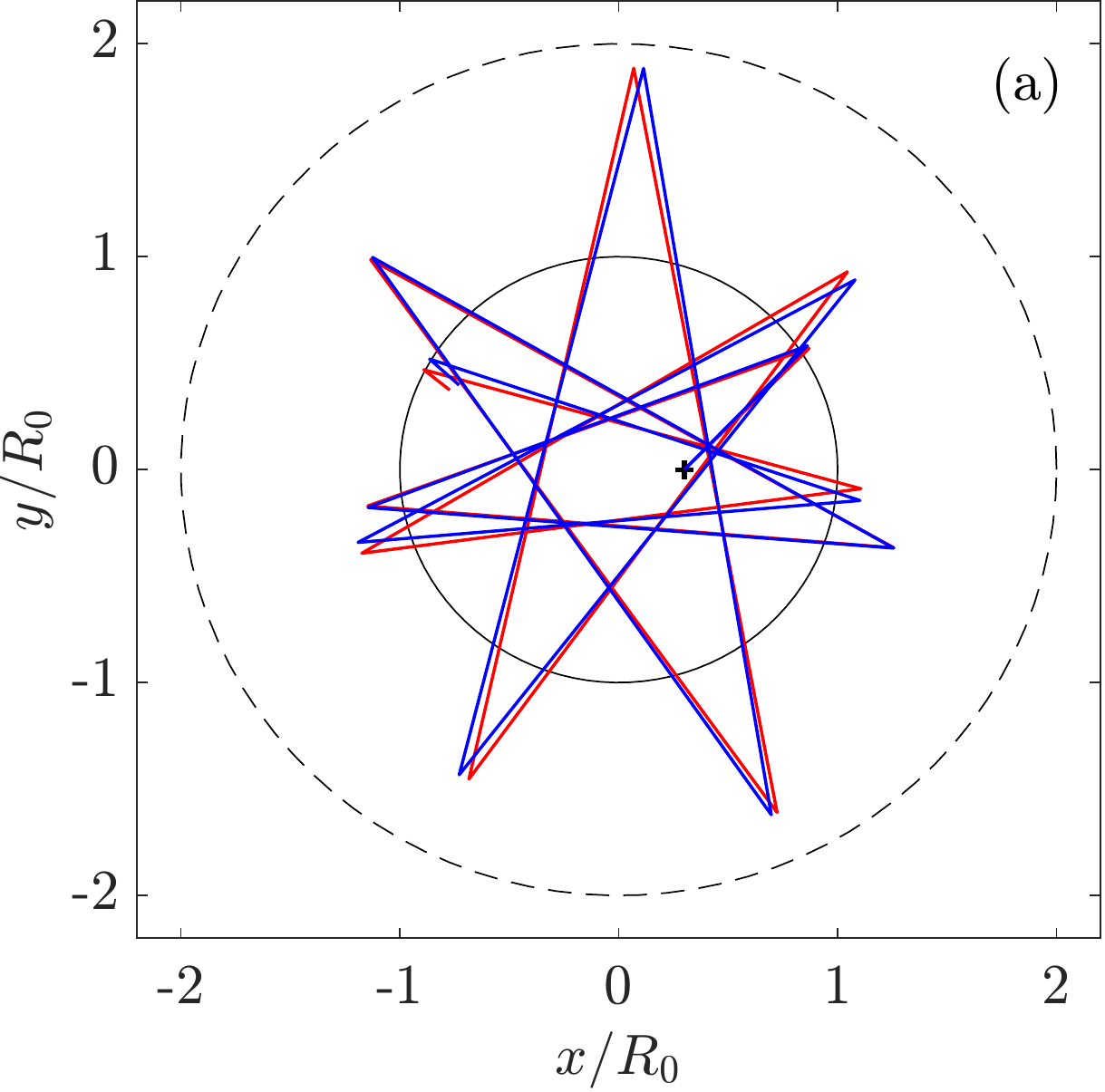} \,\,
\includegraphics[width=.48\linewidth]{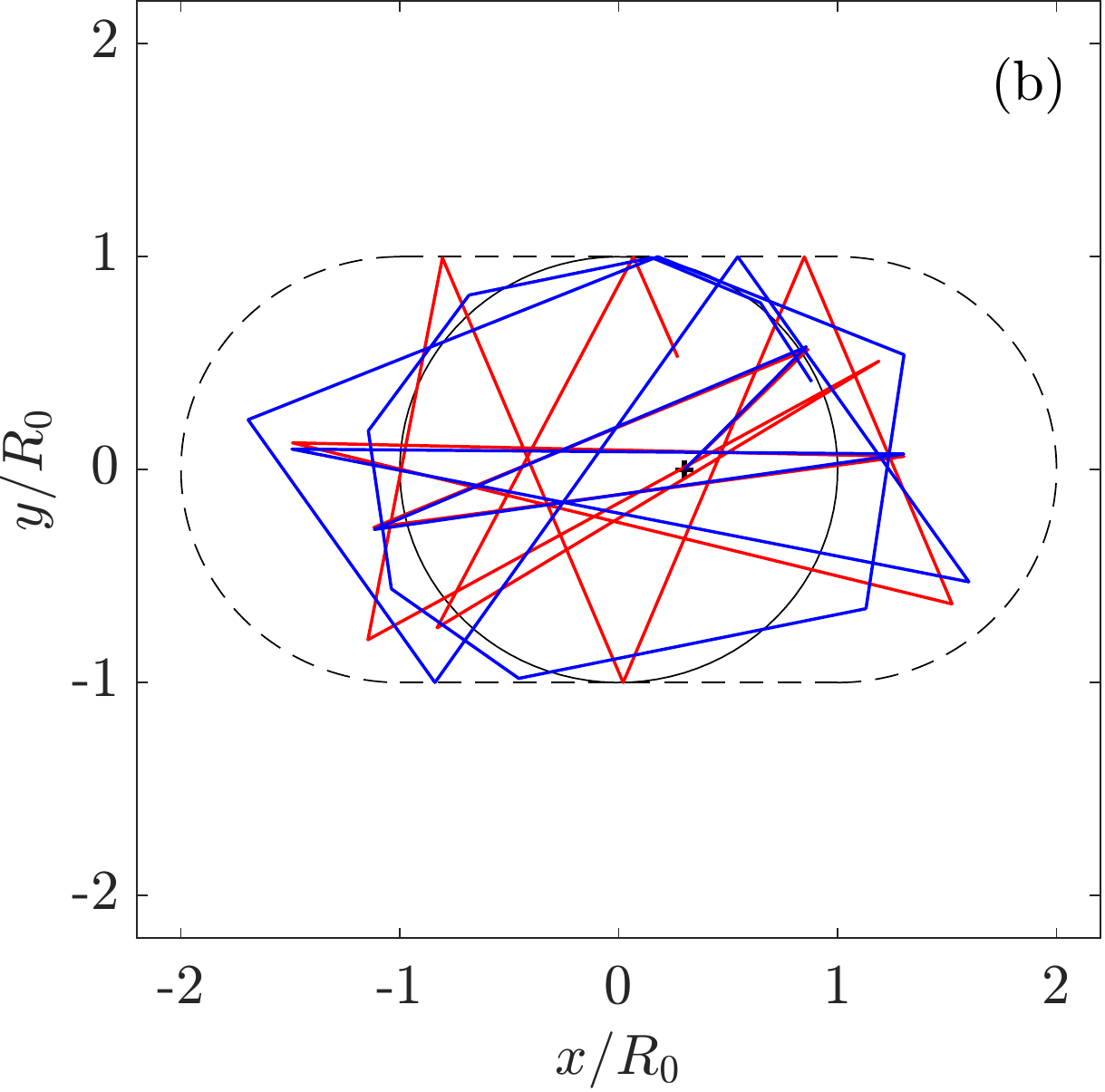}
\caption{Typical trajectories of classical particles at $x_0=0.3 R_0$, $y_0=0$, $v=20u$ (bold plus sign) and a shooting angle of $45^\circ$ (red) and $46^\circ$ (blue) in (a): \textit{System 1} and in (b): \textit{System 2}.}
\label{fig:classTraj}
\end{figure}

Two typical classical trajectories for both systems are shown in Fig.~\ref{fig:classTraj}. The initial conditions of the classical particles are $x_0=0.3 R_0$, $y_0=0$ and $v=20u$; only the shooting angle varies: $45^\circ$ and $46^\circ$. In \textit{System 1} both trajectories stay nearby which is typical for regular systems whereas in \textit{System 2} the trajectories differ strongly after few collisions which is characteristic for chaotic systems. 
 
\begin{figure}
\centering
\includegraphics[width=.48\linewidth]{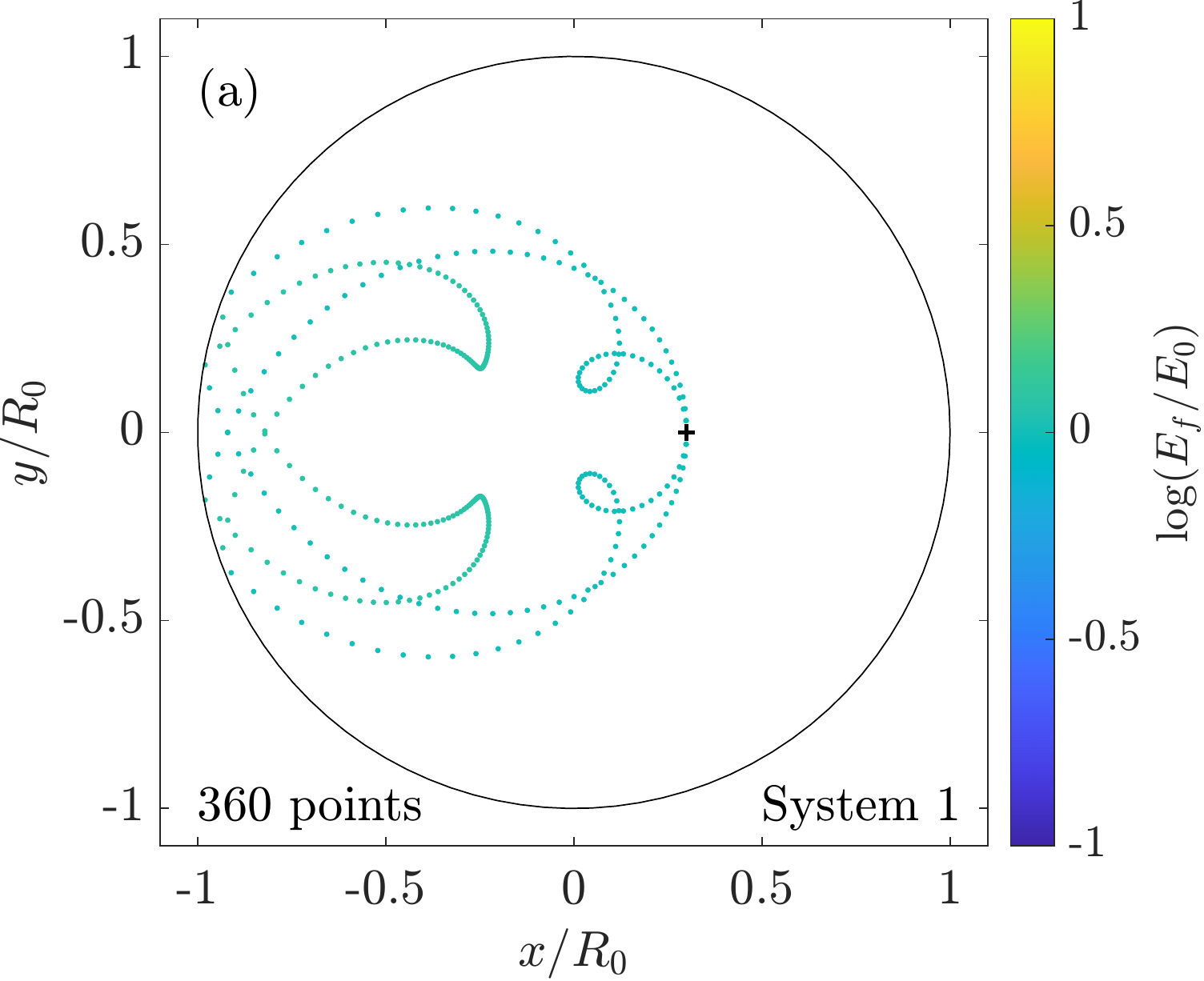} \,\,
\includegraphics[width=.48\linewidth]{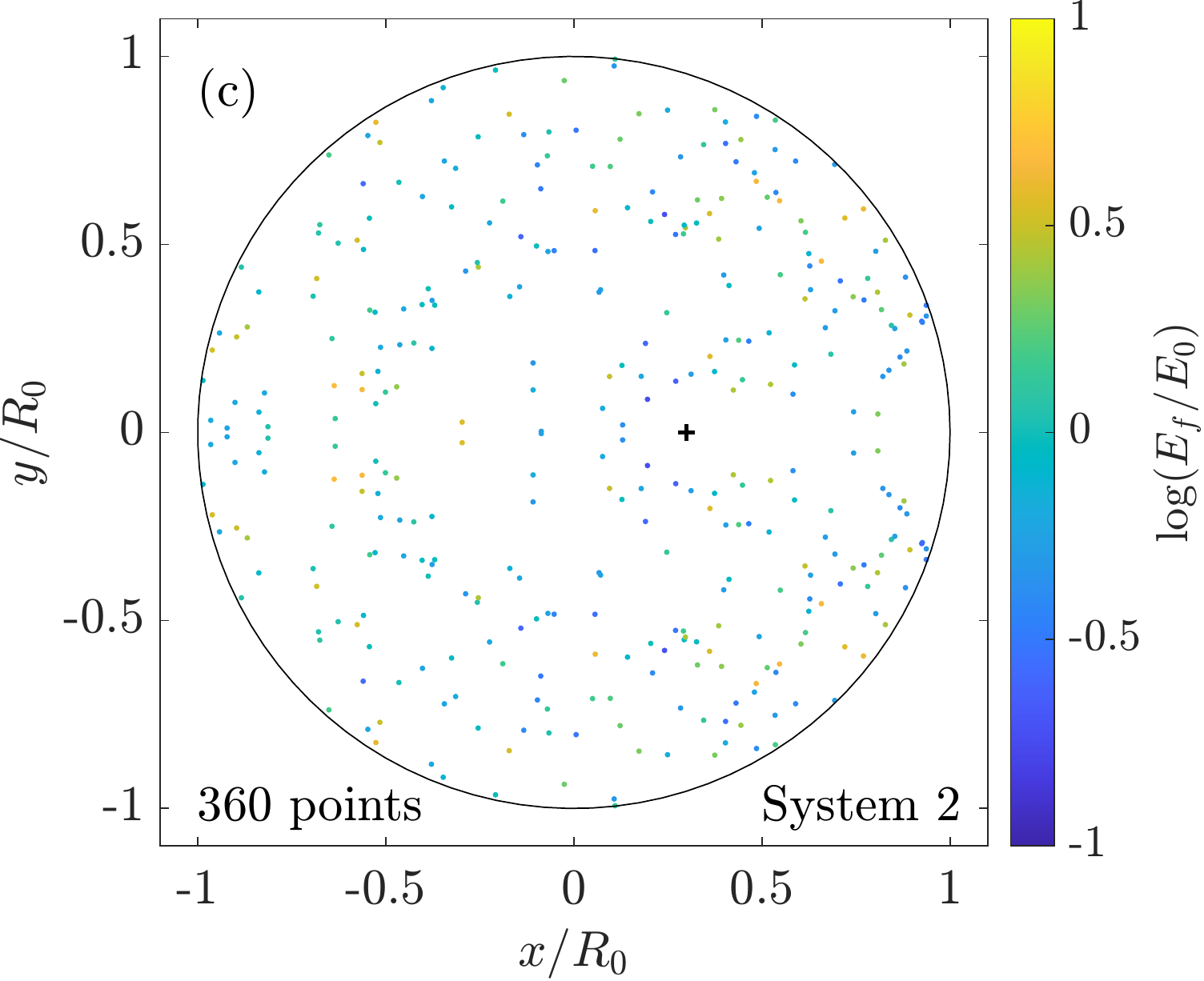} \\
\includegraphics[width=.48\linewidth]{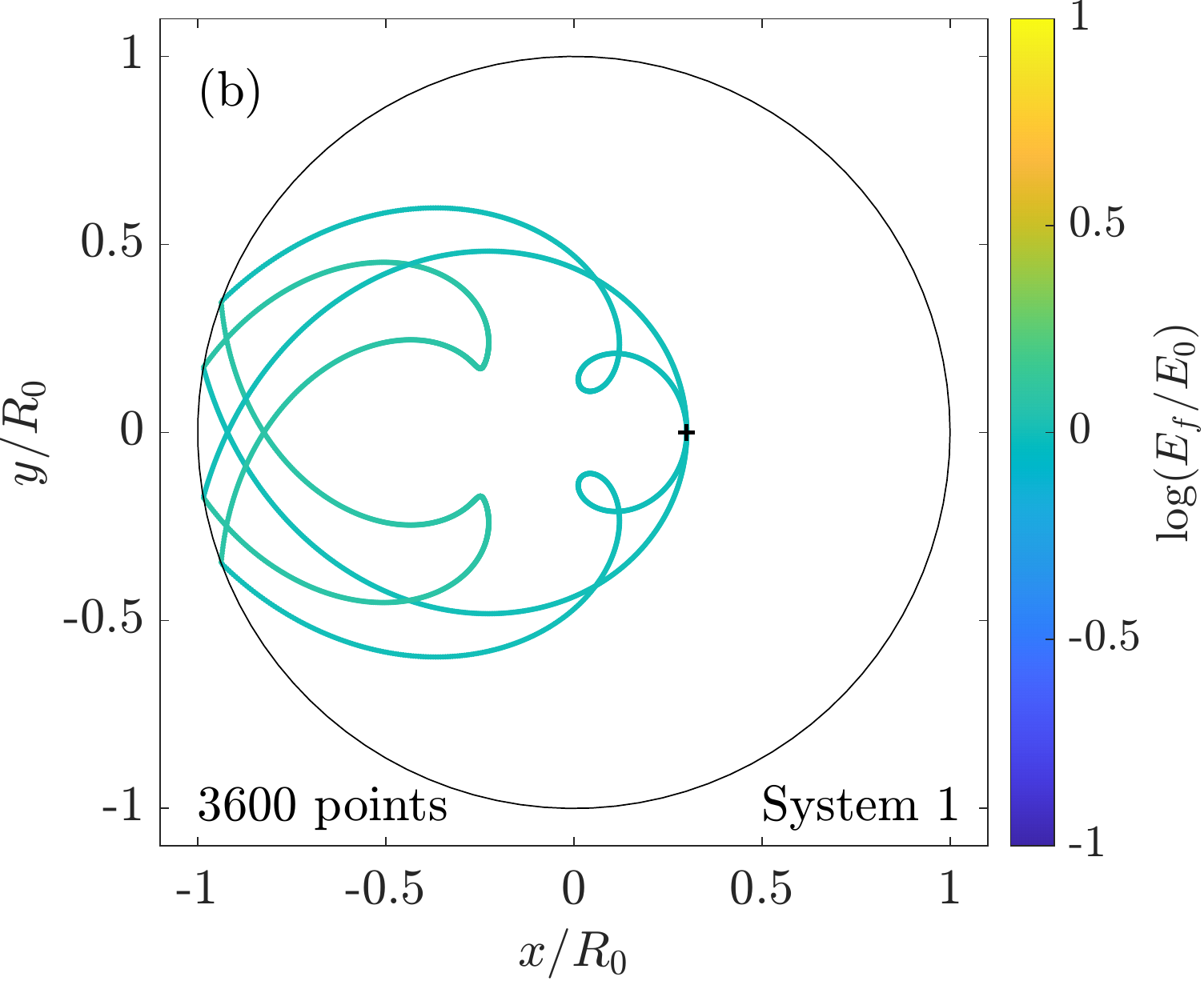} \,\,
\includegraphics[width=.48\linewidth]{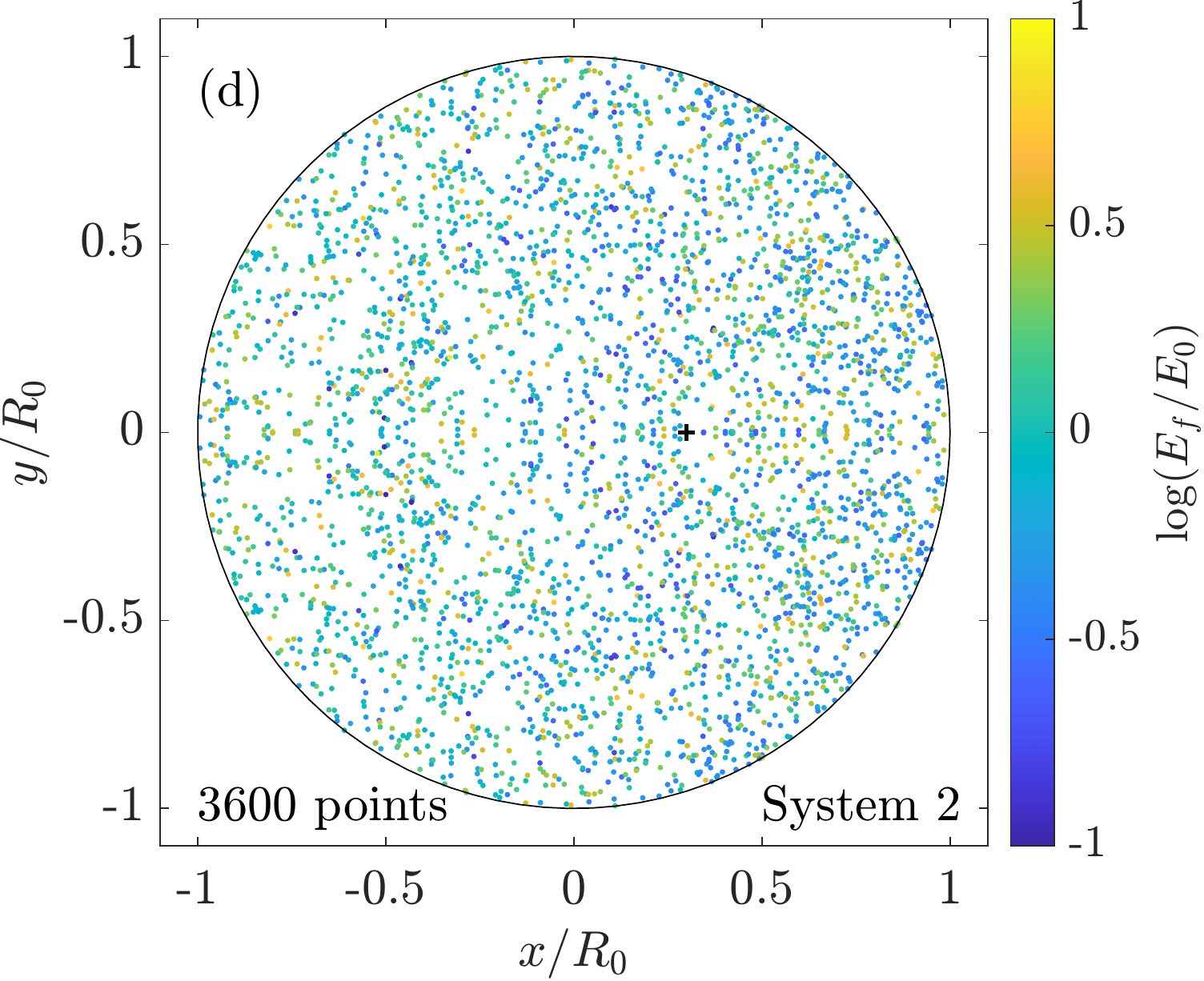} 
\caption{Trajectory end points of classical particles with initial conditions $x_0=0.3 R_0$, $y_0=0$ and $v=20u$ (bold plus sign) at different shooting angles between $0$ and $2\pi$ for \textit{System 1} (a,b) and for \textit{System 2} (c,d). In (a) and (c): $360$, and in (b) and (d): $3600$ equidistant varying shooting angles are used.}
\label{fig:classEnd}
\end{figure}

Using the same initial conditions as in Fig.~\ref{fig:classTraj} but with equidistant shooting angles in the full circle interval $[0,2\pi)$ the end points of all trajectories are plotted in Fig.~\ref{fig:classEnd}. For \textit{System 1} adjacent shooting angles result in nearby end points whereas in \textit{System 2} this behaviour is not occurring even with a more accurate sampling. Also shown are the end energies $E_f$ of these particles related by the initial energy $E_0=\frac{m}{2}v_0^2$. We find that the energy spectrum of \textit{System 2} is broader than that of \textit{System 1}.

\subsection{Quantum mechanical description} \label{sec:quantum}
To describe the quantum dynamics we analyze the evolution of wave functions.
As an initial wave function we start in an eigenstate of the static circular billiard.
The time-independent Schrödinger equation is given by
\begin{align}
E \, \psi &= \left[ - \frac{\hbar^2}{2m} \left( \partial_{\rho}^2 + \frac{1}{\rho} \partial_{\rho} + \frac{\partial_\phi^2}{\rho^2} \right) + V(\rho) \right] \psi  \, , 
\end{align}
where $V(\rho)$ vanishes inside the circle $\rho \leq R_0$ and is infinite otherwise.
The eigenstates are related to the Bessel functions of the first kind, see \cite{Babetal15,AbrSte72},
\begin{align}
 \psi_{n,l}(\rho;R_0) &= \frac{\cal N}{R_0} \, e^{i l \phi} \, J_{l}\left( \frac{\rho}{R_0} j_{n,l} \right) \, .
\end{align}
Here, $j_{n,l}$ denotes the $n$-th zero of the $l$-th Bessel function $J_l$.
The main quantum number $n=1,2,3...$ and the angular momentum quantum number $l=0,1,2...$ define the state of the wave function and are related to the angular momentum by
\begin{align}
 L &= \hbar l \, ,
\end{align}
and to the energy by 
\begin{align}
\label{eq:Enl}
 E_{n,l} &= \frac{\hbar^2}{2mR_0^2} j_{n,l}^2 \, .
\end{align}

For \textit{System 1} we have to solve the time-dependent Schrödinger equation
\begin{align}
i \hbar \partial_t \Psi &= \left[ - \frac{\hbar^2}{2m} \left( \partial^2_{\rho} + \frac{1}{\rho}\partial_{\rho}  + \frac{ \partial^2_{\phi}}{\rho^2}  \right) + V(\rho,t) \right] \Psi \, , 
\end{align}
where $V(\rho,t)$ vanishes inside the circle $\rho \leq R(t)$ and is infinite otherwise. $R(t)$ is given by Eq.~\eqref{eq:Rt}.
This Schrödinger equation can be solved analytically \cite{Babetal15}, the solution reads in the case of expansion $0 \leq t \leq \frac{T}{2}$
\begin{align}
 \Psi_{n,l}(\rho,t) &= {\rm exp}\left( -i\frac{\hbar^2 j_{n,l}^2 t -m^2 u \rho^2 R_0}{2\hbar m R_0 R(t)} \right)  \nonumber \\
 &\quad \times \psi_{n,l}(\rho;R(t)) \, ,
\end{align}
and in the case of contraction $\frac{T}{2} < t \leq T$
\begin{align}
 \Psi_{n,l}(\rho,t) &= {\rm exp}\left(-i \frac{\hbar^2 j_{n,l}^2 \left(t- \frac{T}{2}\right) +m^2 u \rho^2 R_{T/2}}{2\hbar m R_{T/2} R(t)} \right) \nonumber \\
 &\quad \times \psi_{n,l}(\rho;R(t)) \, ,
\end{align}
where $R_{T/2}=R(T/2)$.

The full quantum mechanical problem is solved by three expansions of the initial wave function: first in eigenstates of the time-dependent Schrödinger equation in the expanding case at $t=0$, second by expanding these wave functions at $t=T/2$ in eigenstates of the contracting case at $t=T/2$, and third by expanding these eigenstates at $t=T$ in eigenstates of the time-independent Schrödinger equation.

Unlike \textit{System 1} the potential in \textit{System 2} is not radially symmetric. The time-dependent Schrödinger equation is given by
\begin{align}
i \hbar \partial_t \Psi &= \underbrace{\left[ - \frac{\hbar^2}{2m} \left( \partial^2_{x} + \partial^2_{y} \right) + V(x,y,t) \right]}_{\hat{H}(t)} \Psi \, ,
\end{align}
where $V(x,y,t)$ vanishes inside the stadium and is infinite outside of it.

We solve this Schrödinger equation with the spectral method \cite{FeiFleSte82}. Considering the formal solution
\begin{align}
\Psi(t) &= \hat U(t,0) \Psi(0) \, ,
\end{align}
for short time steps $\Delta t$ we use the time-evolution operator $\hat U(t+\Delta t,t) \approx e^{-i \hat H(t) \Delta t/\hbar}$ to calculate iteratively the wave function in the next time-step
\begin{align}
\Psi(t+\Delta t) &= e^{-i \hat H(t) \Delta t/\hbar}\Psi(t) \, .
\end{align}
At each time-step we split the time evolution operator in three parts
\begin{align}
 e^{-i \hat H \Delta t/\hbar} &\approx e^{i \frac{\hbar}{2m} \frac{\Delta t}{2} (\partial^2_x+\partial^2_y)} e^{-i \hat V \Delta t/\hbar} e^{i \frac{\hbar}{2m} \frac{\Delta t}{2} (\partial^2_x+\partial^2_y)} \, ,
\end{align}
where the derivatives are calculated in the Fourier-space. 

\begin{figure}
\centering
\includegraphics[width=.48\linewidth]{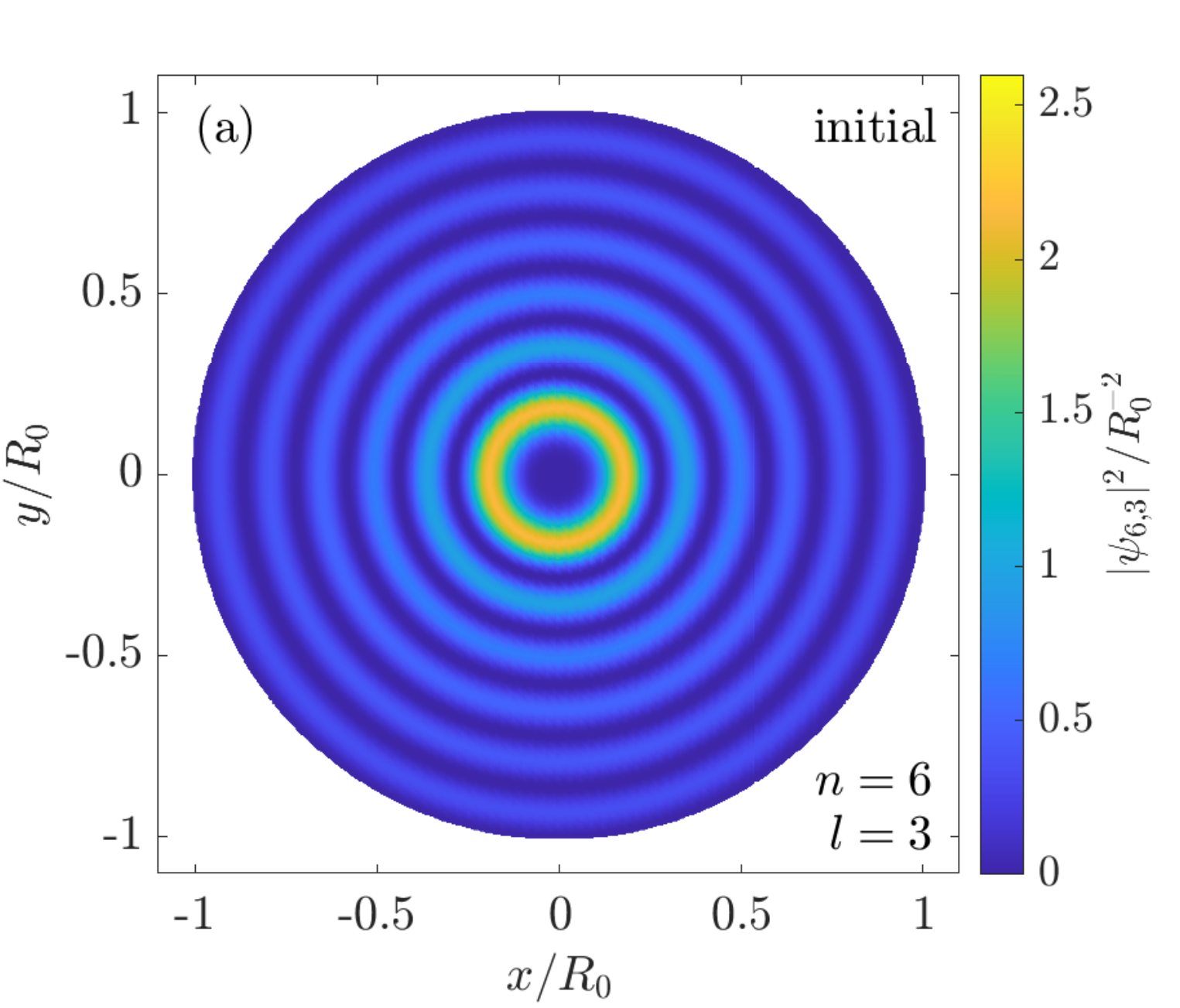} \\
\includegraphics[width=.48\linewidth]{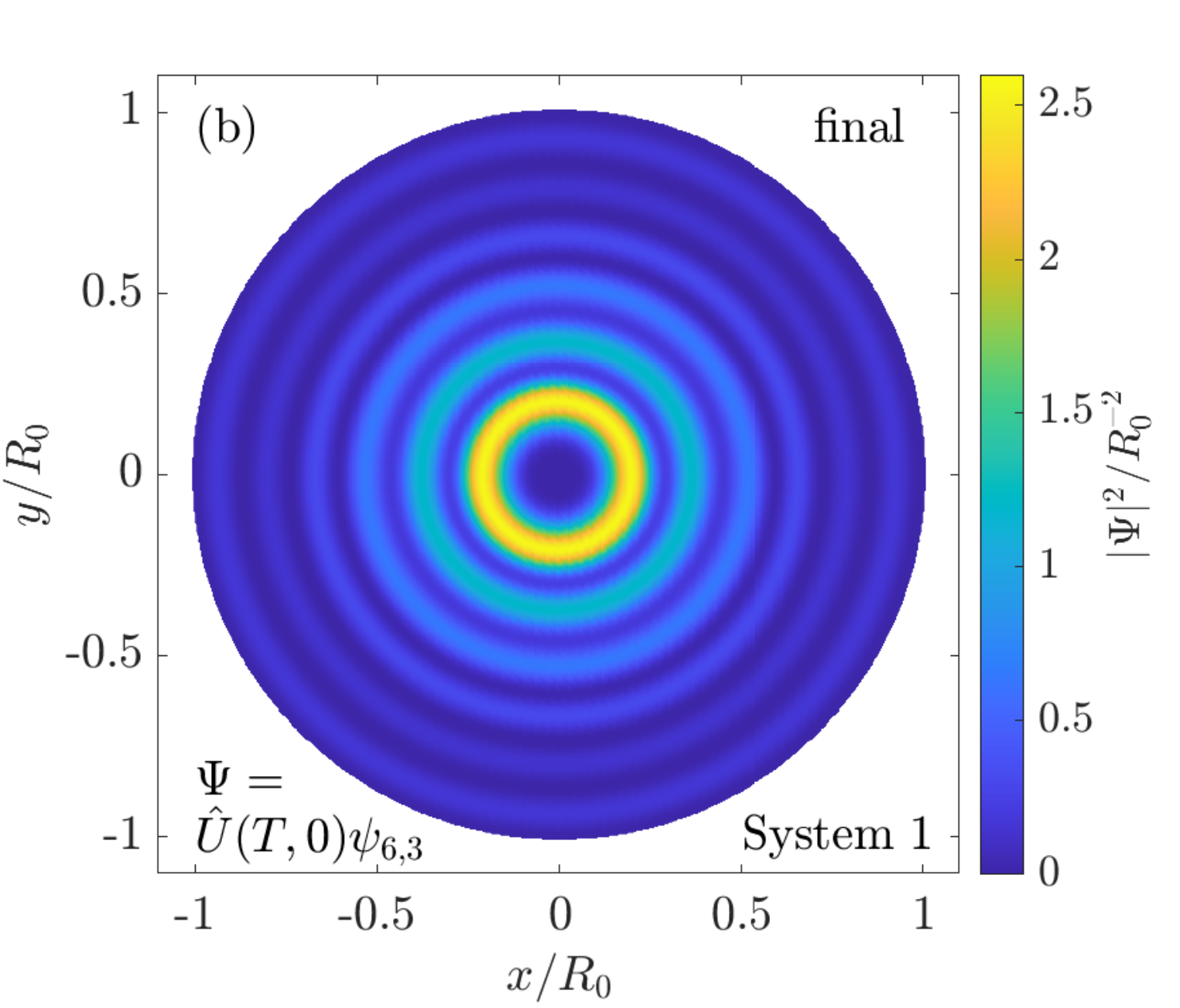} \,\,
\includegraphics[width=.48\linewidth]{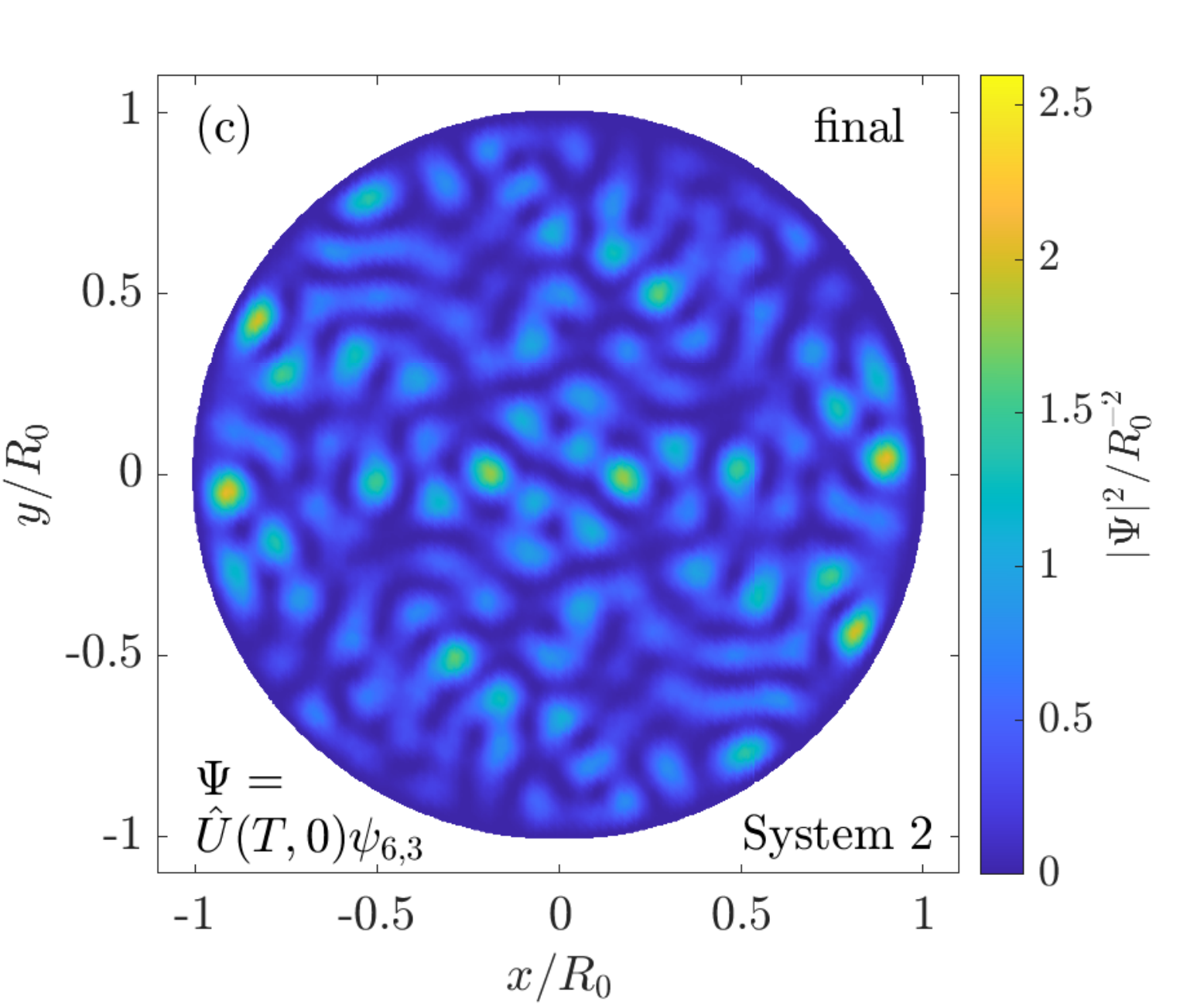}
\caption{Initial wave function for $n=6$ and $l=3$ (a) and its final state in \textit{System 1} (b) or in \textit{System 2} (c). The calculations were performed for Eq.~\eqref{eq:def}.}
\label{fig:QMWF}
\end{figure}

The different evolutions of a quantum mechanical wave function in an eigenstate of the initial circular billiard ($n=6$, $l=3$) are illustrated in Fig.~\ref{fig:QMWF} in which the squared moduli of the wave functions are plotted. Note, that in the following all quantum mechanical calculations are performed for

\begin{align}
\label{eq:def}\frac{\hbar T}{2mR_0^2}=1~\Rightarrow ~ j_{n,l} = \frac{v_0}{u} \, .
\end{align}
So the zeros of the Bessel functions are comparable to the classical velocity $v_0$. Because of angular momentum conservation in \textit{System 1}, the radial symmetry of the initial wave function is conserved. This is different to the final wave function after evolution in \textit{System 2}. There the angular momentum is not conserved and therefore the final wave function is not radial symmetric. But because of the point symmetry of the system the final wave function is point symmetric.

\section{Results} \label{sec:results}

%


\subsection{Transition probabilities} \label{sec:trans}

Both dynamical billiard systems start and end in a circular billiard with radius $R_0$. 

In classical mechanics a given initial position ${\bf r}_0$ and velocity ${\bf v}_0$ of a particle determine the final position ${\bf r}_f$ and velocity ${\bf v}_f$. But the knowledge of starting energy $E_0$ and angular momentum $L_0$ is not enough to determine the final energy $E$ and angular momentum $L$. 
For statistics we express transitions via the joint conditional probability density $p(E,L|E_0,L_0)$ which gives the joint probability density of the final energy $E$ and angular momentum $L$ under the condition to start with the energy $E_0$ and angular momentum $L_0$. This probability density is calculated by numerical simulation of $10^5$ classical particles as discussed in App.~\ref{sec:AppA}.

In the quantum case we represent the wave functions at the end as a superposition of eigenstates of the circular billiard with radius $R_0$. So the probability for transitions from state $(n_0,l_0)$ to $(n,l)$ is given by
\begin{align}
 P(n,l|n_0,l_0) &= 
\left|\left\langle \psi_{n,l} \left| \hat{U}(T,0) \right| \psi_{n_0,l_0} \right\rangle \right|^2 \, . 
\end{align}

For a comparison of classical and quantum mechanical results we introduce the cumulative conditional probability of the final energy $F(E|E_0)$ which is classical defined by 
\begin{align}
 F^{\rm cl}(E|E_0) &= \int\limits_{0}^{\infty} dE' \, \theta\left(E - E' \right) p(E'|E_0) \, , \\
 p(E'|E_0) &= \int\limits_{-\infty}^{\infty} dL\int\limits_{-\infty}^{\infty} dL_0 \, p(E',L|E_0,L_0) \, .
\end{align}
In the quantum case it is defined by
\begin{align}
 F^{\rm qm}(E|E_{n_0,l_0}) &= \sum_{n,l} \theta\left( E - E_{n,l} \right) P(n,l|n_0,l_0) \, ,
\end{align}
with the Heaviside function $\theta(x)=\left\{\begin{smallmatrix} 0 & x<0 \\ 1 & x \geq 0 \end{smallmatrix}\right.$.

\begin{figure}
\centering
\includegraphics[width=.48\linewidth]{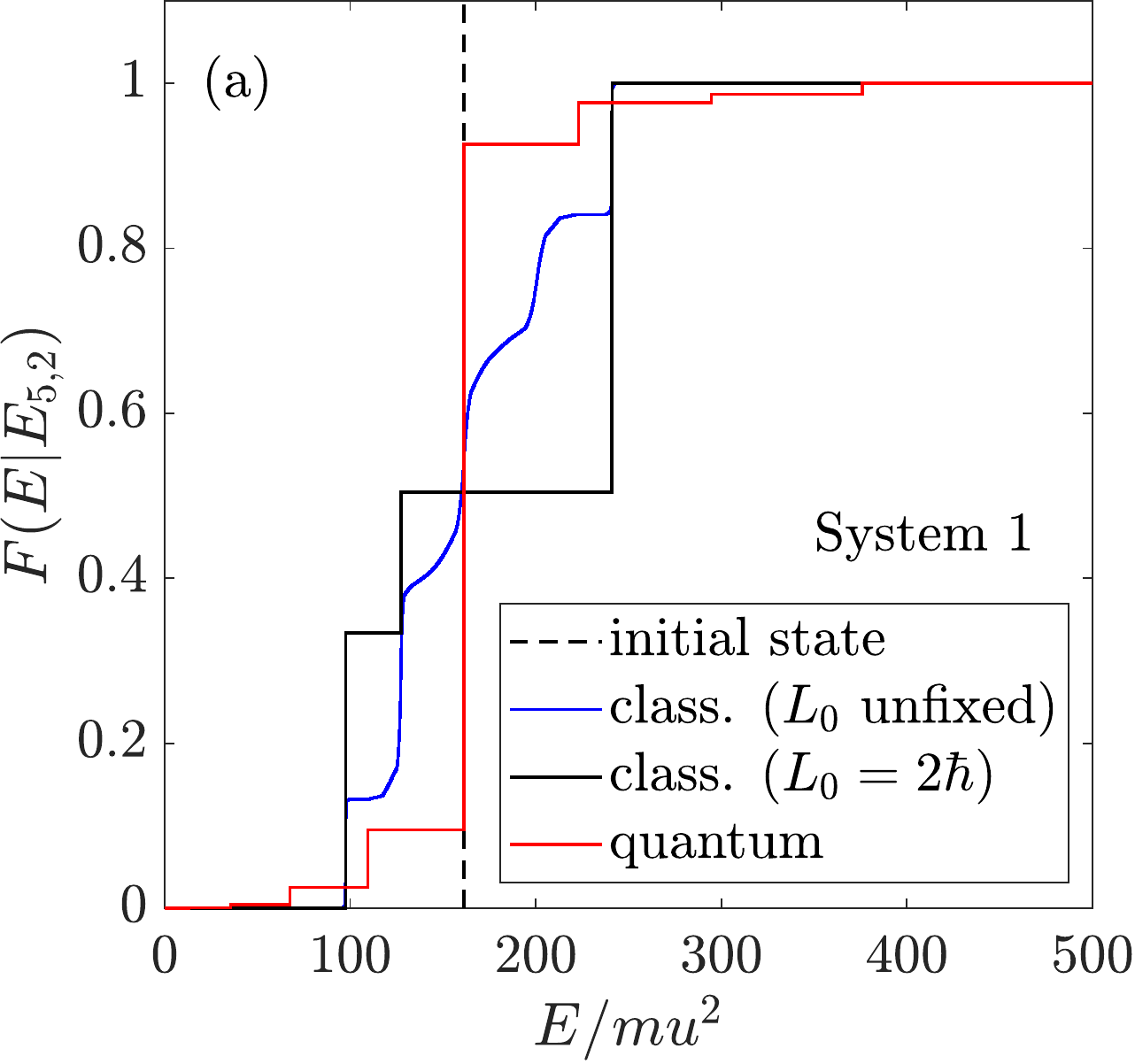} \,\,
\includegraphics[width=.48\linewidth]{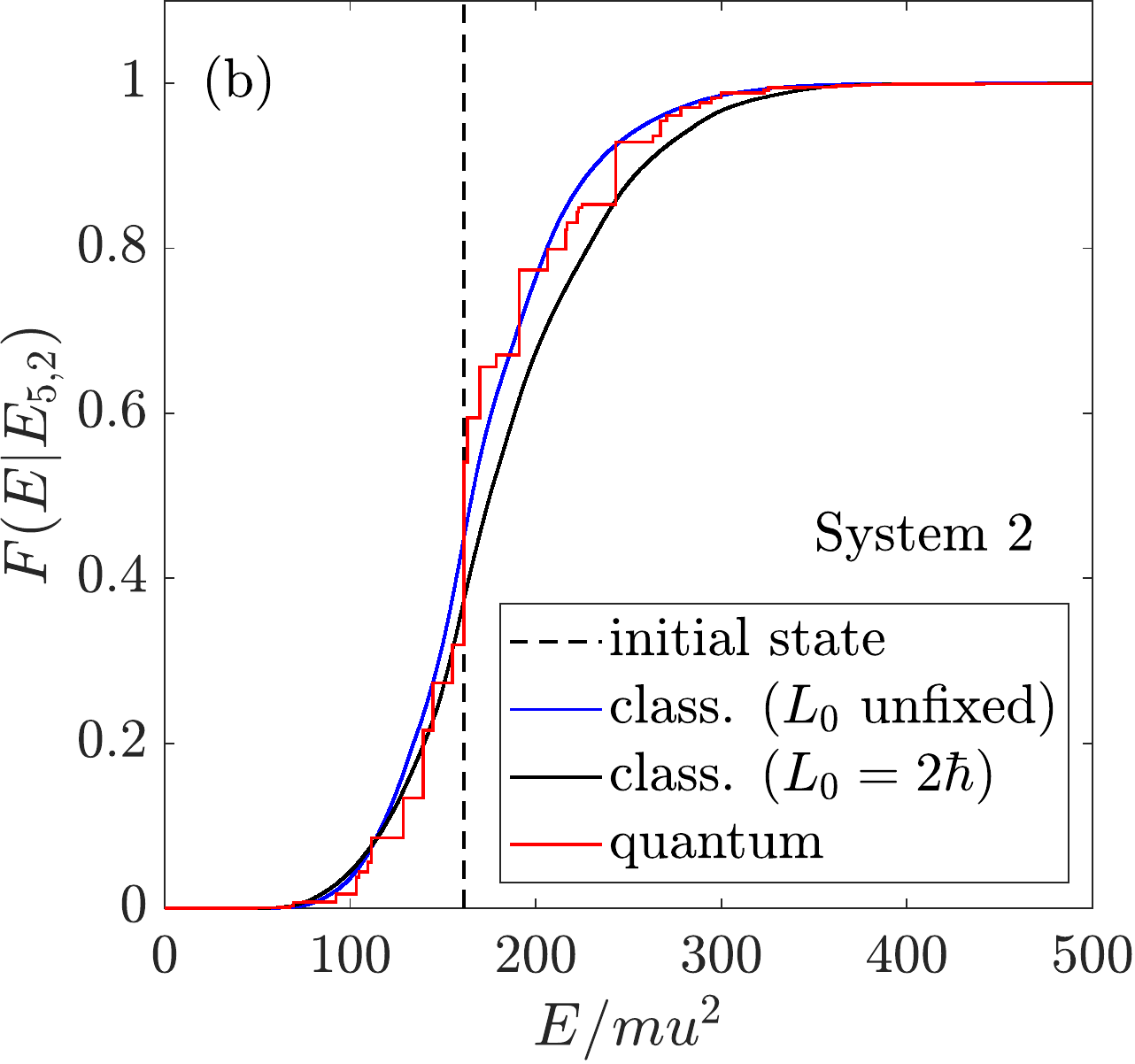}
\caption{Cumulative conditional probability for (a): \textit{System 1} and (b): \textit{System 2} with Eq.~\eqref{eq:def}. Classical calculations were performed for $10^5$ particles with an initial kinetic energy of $E_{5,2}$ regardless of the angular momentum (blue) and with explicit consideration of $L=\hbar l_0$ (solid black). In the quantum case (red) we start with a wave function in the eigenstate $n_0=5$ and $l_0=2$. The initial energy is illustrated by a dashed black line.}
\label{fig:FEE}
\end{figure}

For the initial conditions $n_0=5$ and $l_0=2$ which are related to the energy $E_0 \approx 160mu^2$ ($v_0 \approx 18u$) the results are shown in Fig.~\ref{fig:FEE}. 
In the case of a fixed classical angular momentum $L_0 = 2\hbar$ we find in \textit{System 1} a step-like function. Along each jump the number $n_e$ of energy losing collisions in the expanding phase is equal for all trajectories as well as the number $n_k$ of energy gaining collisions in the contracting phase. In the first visible jump at $E \approx 100mu^2$ these numbers are $n_e=5$ and $n_k=3$, in the second at $E \approx 130mu^2$ they are $n_e=5$ and $n_k=4$, and in the last jump at $E \approx 240mu^2$ they are $n_e=4$ and $n_k=6$. Other combinations are not observed. Because of the small angular momentum the particle velocity loss/gain per collision is almost equal. Note, only in the special case $L_0=0$ the particle velocity change is exact $2u$ per collision which is the well known result by particle collisions on moving walls \cite{LanLifI17}. With increasing values of $L_0$ the energy loss/gain per collision fluctuates more. This leads to sigmoidal steps in the results at arbitrary (unfixed) angular momenta which is a superposition of all possible angular momenta. Nevertheless, the dominant jumps at lower $L_0$ stay visible. 
The classical results in \textit{System 2} show a more continuous distribution for a fixed angular momentum as well as for the unfixed angular momenta which is a consequence of the chaotic behaviour. Indeed each collision can be predicted analytically for given initial conditions but small changes in these initial conditions will change dramatically the trajectory and the final conditions.

In \textit{System 1} the quantum mechanical result for the cumulative conditional probability show a step-function which is the consequence of countable many transitions. Because of angular momentum conservation only transitions with $l=l_0$ are allowed. In contrast to this the quantum mechanical calculations in \textit{System 2} show a quasi-continuous distribution which is a consequence of the larger number of possible transitions. While in \textit{System 1} the energy gap increases in \textit{System 2} it does not.
Independent of the considered system we conclude that classically forbidden energy transitions are unlikely in the quantum case. 

%
%
%

\subsection{Work distribution} \label{sec:PW}

Both systems are considered initially in thermal equilibrium with a bath at inverse temperature $\beta$ and are decoupled at $t=0$ up to $t=T$. 

We first consider the classical case. Based on the first law of thermodynamics the work is related to the energy difference $W=E-E_0$ and the cumulative work distribution $F(W)$ is given by
\begin{align}
 F^{\rm cl}(W) &= \frac{1}{Z^{\rm cl}} \int\limits_{0}^{\infty} dE \int\limits_{0}^{\infty} dE_0 \, e^{-\beta E_0} \nonumber \\ & \quad \times \theta\left( E_0 - E + W \right) p(E|E_0) \, ,
\end{align}
with the normalization $Z^{\rm cl} = \int_{0}^{\infty} dE_0 e^{-\beta E_0}= \beta^{-1}$.

In calculation of the work in quantum systems we use the two projective measurement method which fulfills the Jarzynski equation $\langle e^{-\beta W} \rangle = 1$ \cite{Jar97, EngNol07}:
First, we measure the energy at the beginning $t=0$. Hence we start in an eigenstate of the static circular billiard. After the second measurement at $t=T$ we end in an eigenstate of the same static circular billiard with probability $P(n,l|n_0,l_0)$. The cumulative work distribution $F(W)$ is given by
\begin{align}
 F^{\rm qm}(W) &= \frac{1}{Z^{\rm qm}} \sum_{n,l,n_0,l_0} e^{-\beta E_{n_0,l_0}}  \nonumber \\ &\quad  \times \theta\left( E_{n_0,l_0} - E_{n,l} + W \right) P(n,l|n_0,l_0) \, ,
\end{align}
with the partition function $Z^{\rm qm}=\sum_{n_0,l_0} e^{-\beta E_{n_0,l_0}}$. 

For the following results we have verified the Jarzinsky equation $\langle e^{-\beta W} \rangle = 1$, in the classical as well as in the quantum mechanical case. This is a consistency check for our numerical calculations.

\begin{figure}
\centering
\includegraphics[width=.48\linewidth]{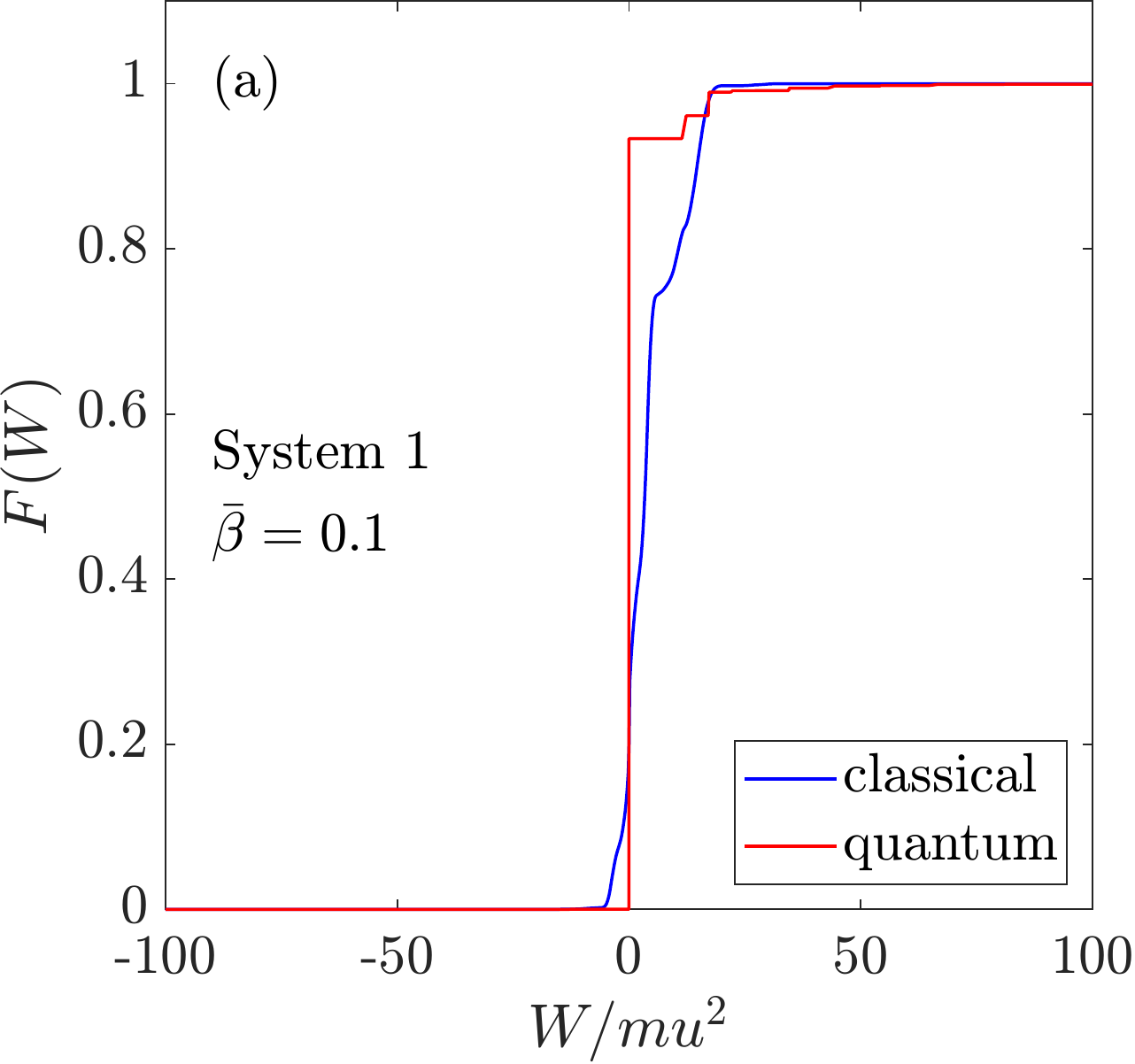} \,\,
\includegraphics[width=.48\linewidth]{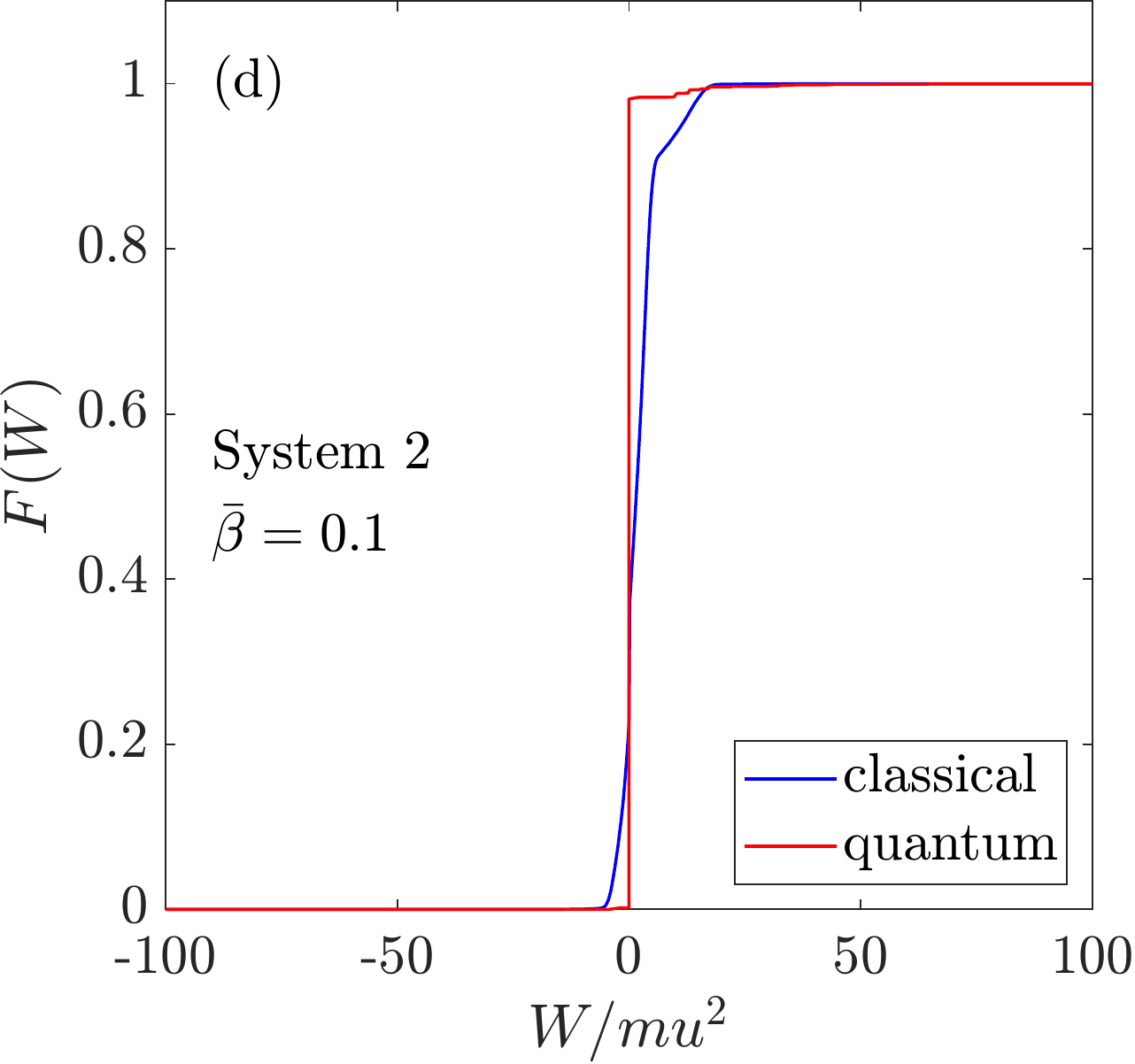} \\
\includegraphics[width=.48\linewidth]{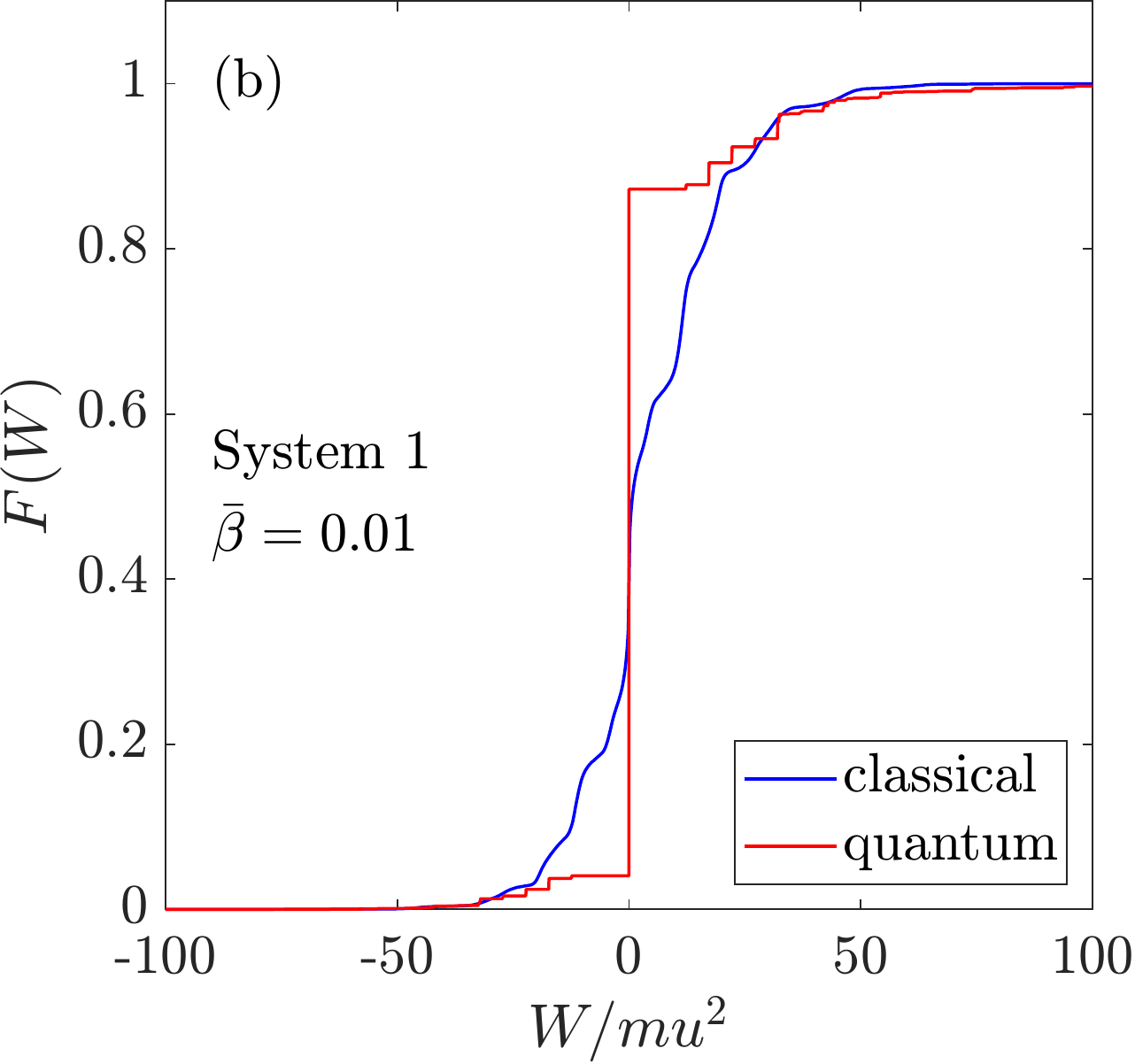} \,\,
\includegraphics[width=.48\linewidth]{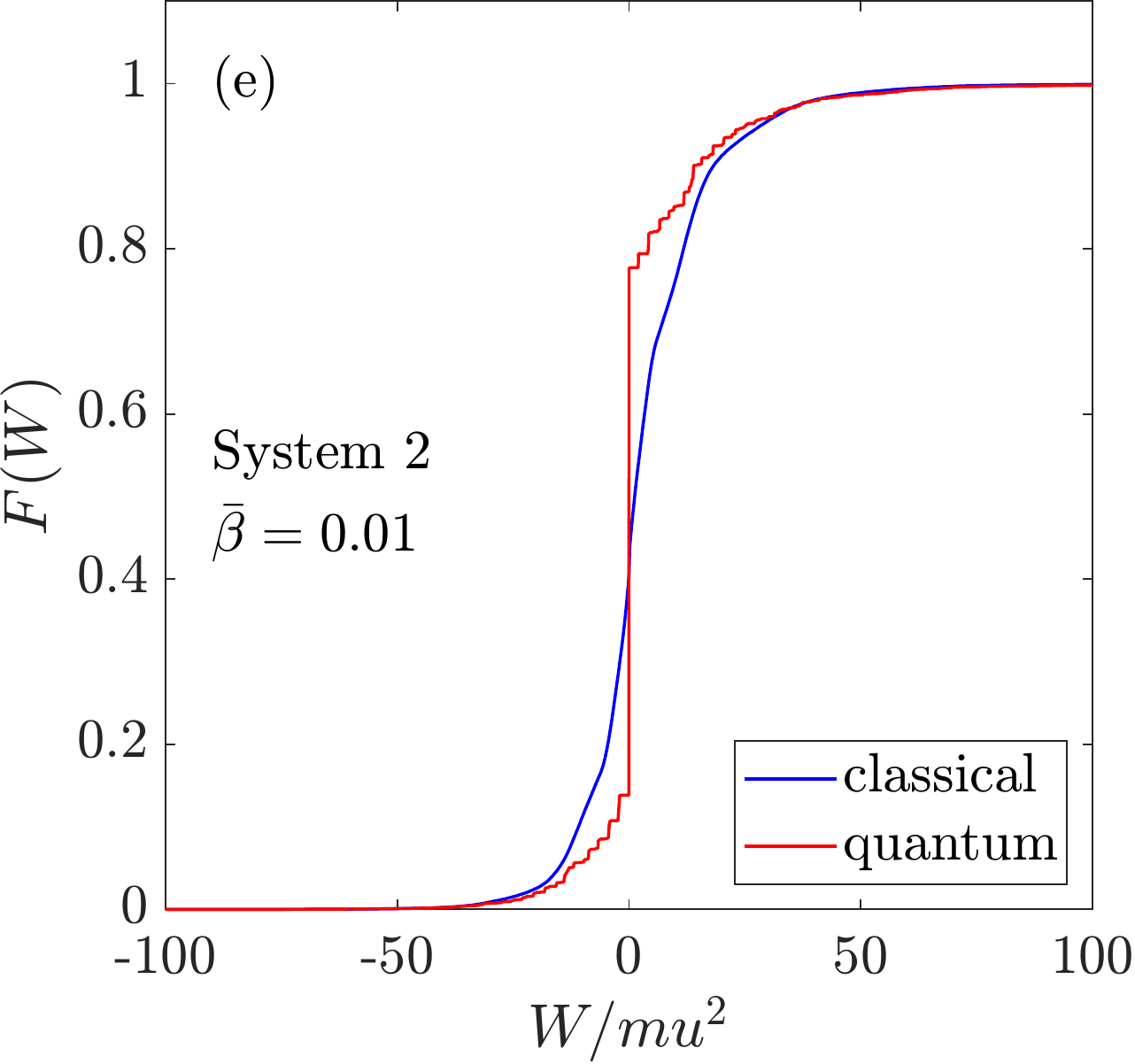} \\
\includegraphics[width=.48\linewidth]{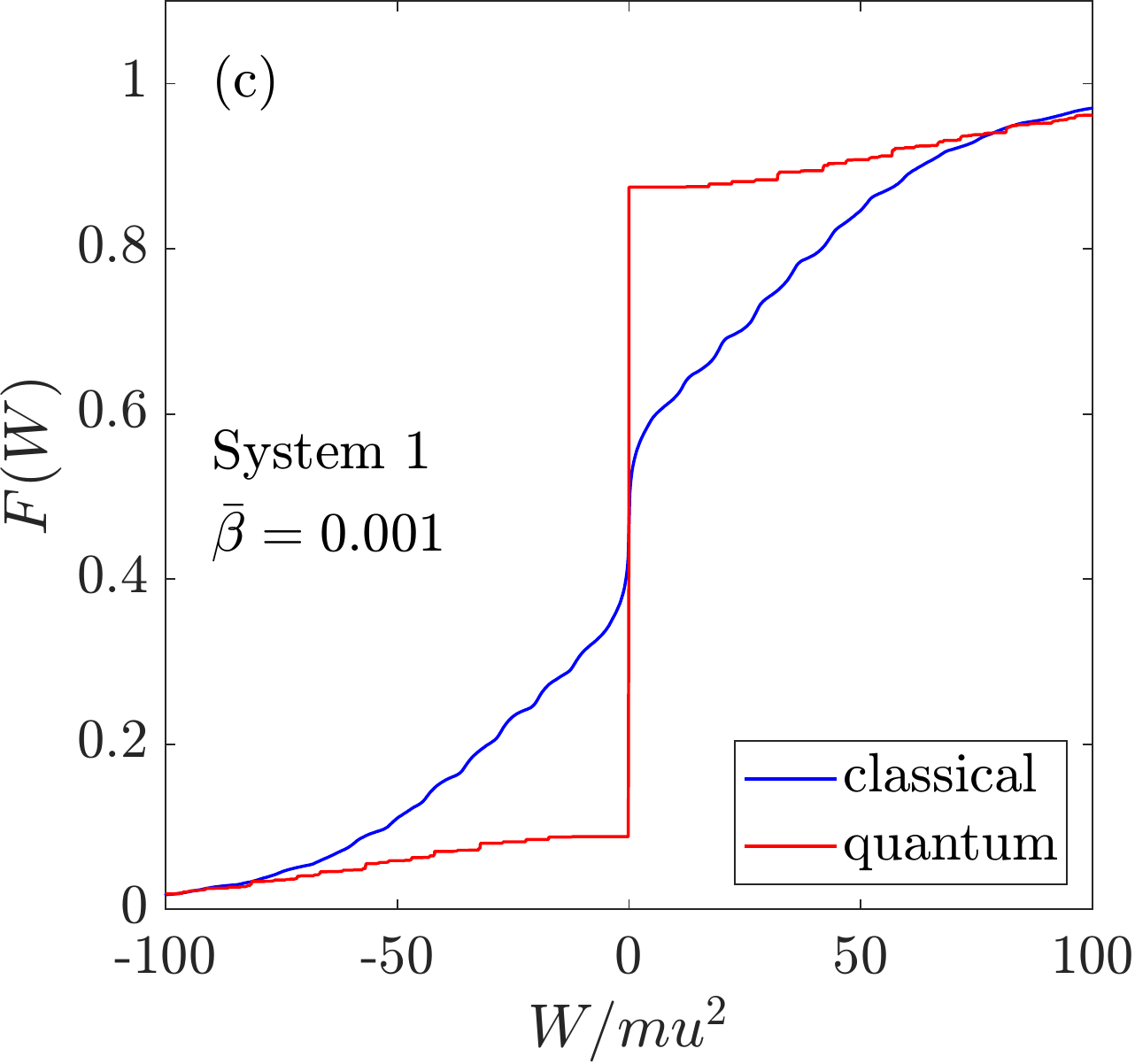} \,\,
\includegraphics[width=.48\linewidth]{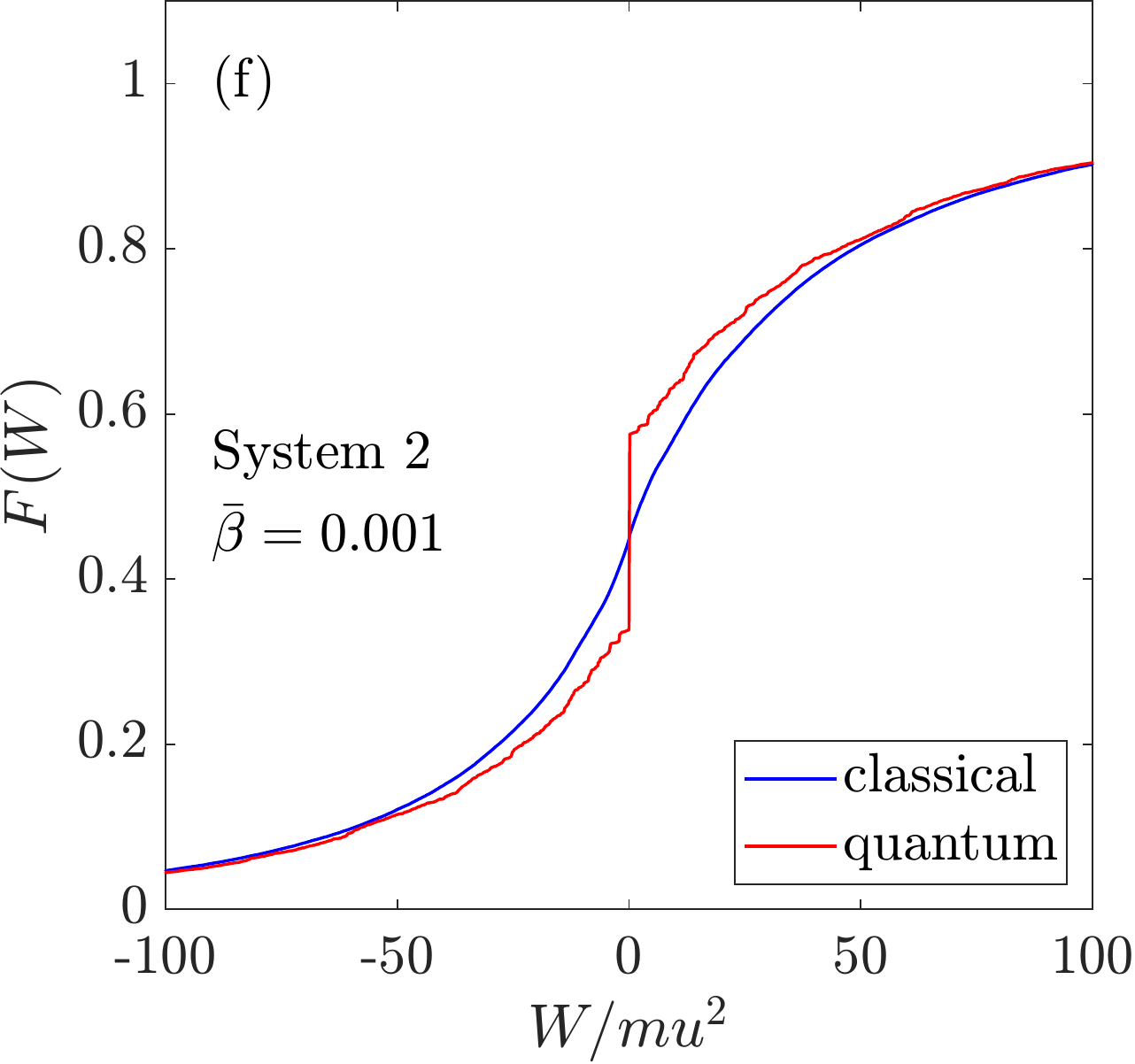} 
\caption{Classical (blue) and quantum mechanical (red) cumulative work distribution at different temperatures $1/\bar{\beta}$ for (a)--(c): \textit{System 1} and for (d)--(f): \textit{System 2}. The quantum mechanical calculations were performed for Eq.~\eqref{eq:def}.}
\label{fig:FW}
\end{figure}

The results for the cumulative work distribution for different temperatures represented by the dimensionless quantity
\begin{align}
 \bar{\beta} &\stackrel{\phantom{\rm a.s.}}{=} \frac{m R_0^2}{T^2}\beta  \label{eq:barbeta} \, ,
\end{align}
are shown in Fig.~\ref{fig:FW}. 
In all figures the classical calculations were performed for $10^5$ particles. For \textit{System 1} we recognize the sigmoidal steps from Fig.~\ref{fig:FEE}(a), for \textit{System 2} the continuous distribution from Fig.~\ref{fig:FEE}(b). As expected in both systems, at higher temperatures (small values of $\bar\beta$) the width of the distribution gets broadened and the jump at $W=0$ decreases. These effects are also visible in the quantum mechanical results. Moreover, for higher temperatures classical and quantum mechanical calculations converge to each other, especially visible in \textit{System 2}. But note, this convergence is limited by the allowed transitions, explained in the following paragraph. 

Because of angular momentum conservation in \textit{System 1} the minimal positive value of work is limited by $E_{2,0}-E_{1,0}$, cf. Eq. (\ref{eq:Enl}). This is different in \textit{System 2} in which transitions with angular momentum change are allowed, too. These transitions might be closer than $E_{2,0}-E_{1,0}$. For example when taking into account the $1000$ lowest states we find $2823$ possible transitions inside the work interval $(0, E_{2,0}-E_{1,0})$ in \textit{System 2} whereas we find no transitions in \textit{System 1}.

%

\subsection{Probability of no energy change} \label{sec:PW0}

In both systems there is a non-zero probability that the processes occur without performing mechanical work in the quantum as well as in the classical case, see Fig.~\ref{fig:FW}. 

In the classical case of \textit{System 1} any collisions yield an energy change except for the occasional one. The probability of zero collisions is derived in App.~\ref{sec:AppB}
\begin{align}
 P_1^{\rm cl}(W=0) &= 1-e^{-\bar{\beta}}\left[ I_0(\bar{\beta})+I_1(\bar{\beta}) \right] \, , \label{eq:P1W0ana}
\end{align}
where $I_0$ and $I_1$ are modified Bessel's functions of the first kind and $\bar{\beta}$ is defined in Eq.~\eqref{eq:barbeta}.

In contrast to this in \textit{System 2} exists a non-zero probability to perform collisions only with the horizontal walls of the stadium which conserve the energy. So the condition of zero collisions is sufficient but not necessary for $W=0$. So we expect $P_1^{\rm cl}(W=0) \leq P_2^{\rm cl}(W=0)$ as a lower limit. Also an upper limit has been derived 
\begin{widetext}
\begin{align}
 P_2^{\rm cl}(W=0) &\leq P_1^{\rm cl}(W=0) + \sum_{k=1}^{\infty}  \frac{2\bar{\beta}}{\pi} \int\limits_{2k-2}^{2k} dx \, x e^{-\frac{\bar{\beta}}{2}x^2} \int\limits_{2k-1}^{1+x} dy \, f(x,y) \cdot \arcsin\sqrt{\frac{16k^2-(y^2-1-4k^2)^2}{16k^2y^2}} \, , \label{eq:P2W0ana} \\
 f(x,y) &\stackrel{\phantom{\rm a.s.}}{=} y - \frac{2y}{\pi} \arctan\left[ \frac{-1+y^2+x^2}{\sqrt{-y^4-(1-x)^2+2y^2\left(1+x^2\right)}} \right] \, .
\end{align}
\end{widetext}
For details see also App.~\ref{sec:AppB}.

Because of non-degenerated eigenvalues in the quantum case a vanishing work is a consequence of self-transitions. So the probability $P(W=0)$ is related to the trace of the transition probabilities 
\begin{align}
 P^{\rm qm}(W=0) = \frac{1}{Z^{\rm qm}} \sum_{n_0,l_0} e^{-\beta E_{n_0,l_0}} P(n_0,l_0|n_0,l_0) \, . \label{eq:PW0qm}
\end{align}
As a consequence the angular momentum conservation is a necessary condition.

\begin{figure}
\centering
\includegraphics[width=.48\linewidth]{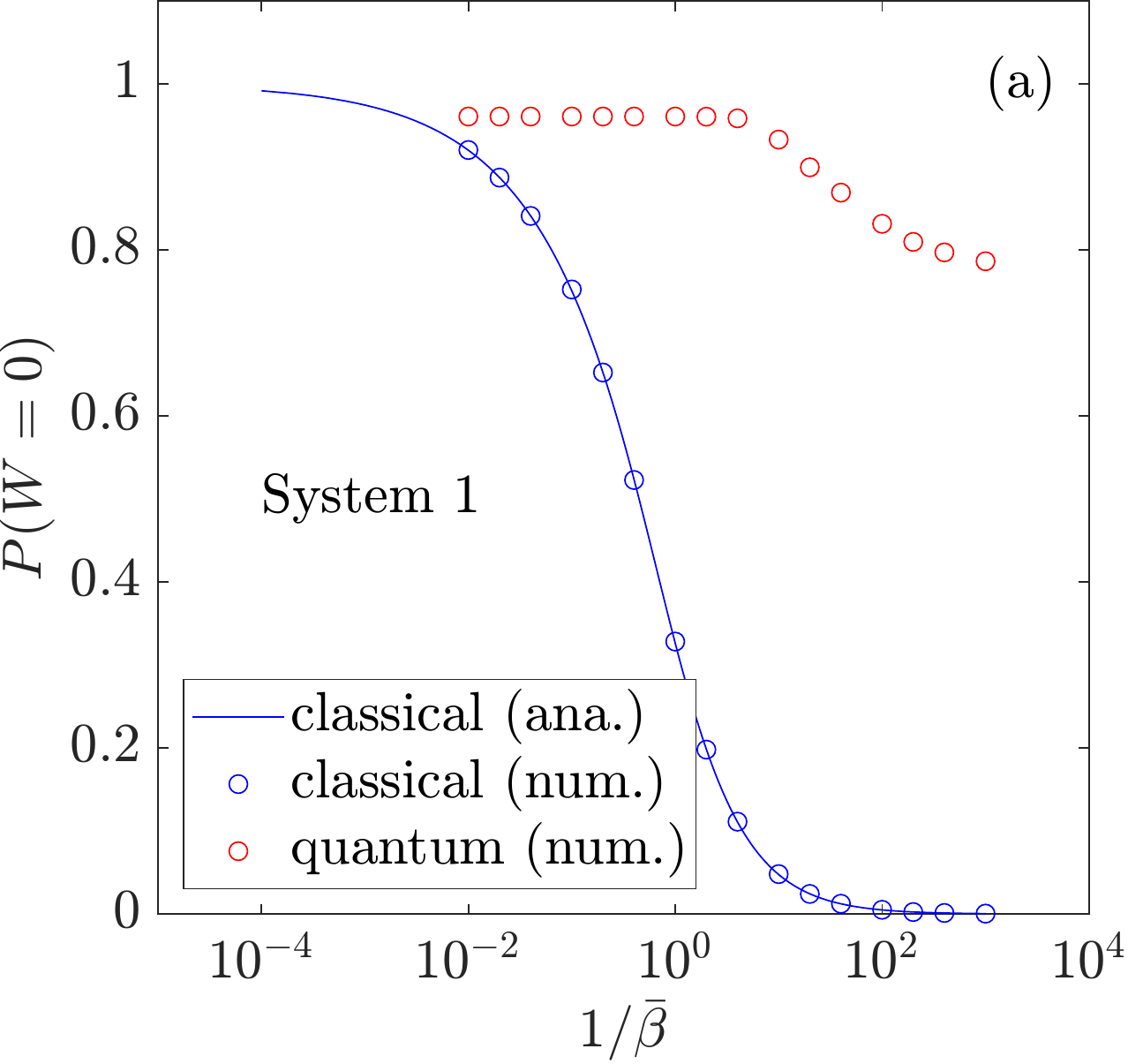} \,\,
\includegraphics[width=.48\linewidth]{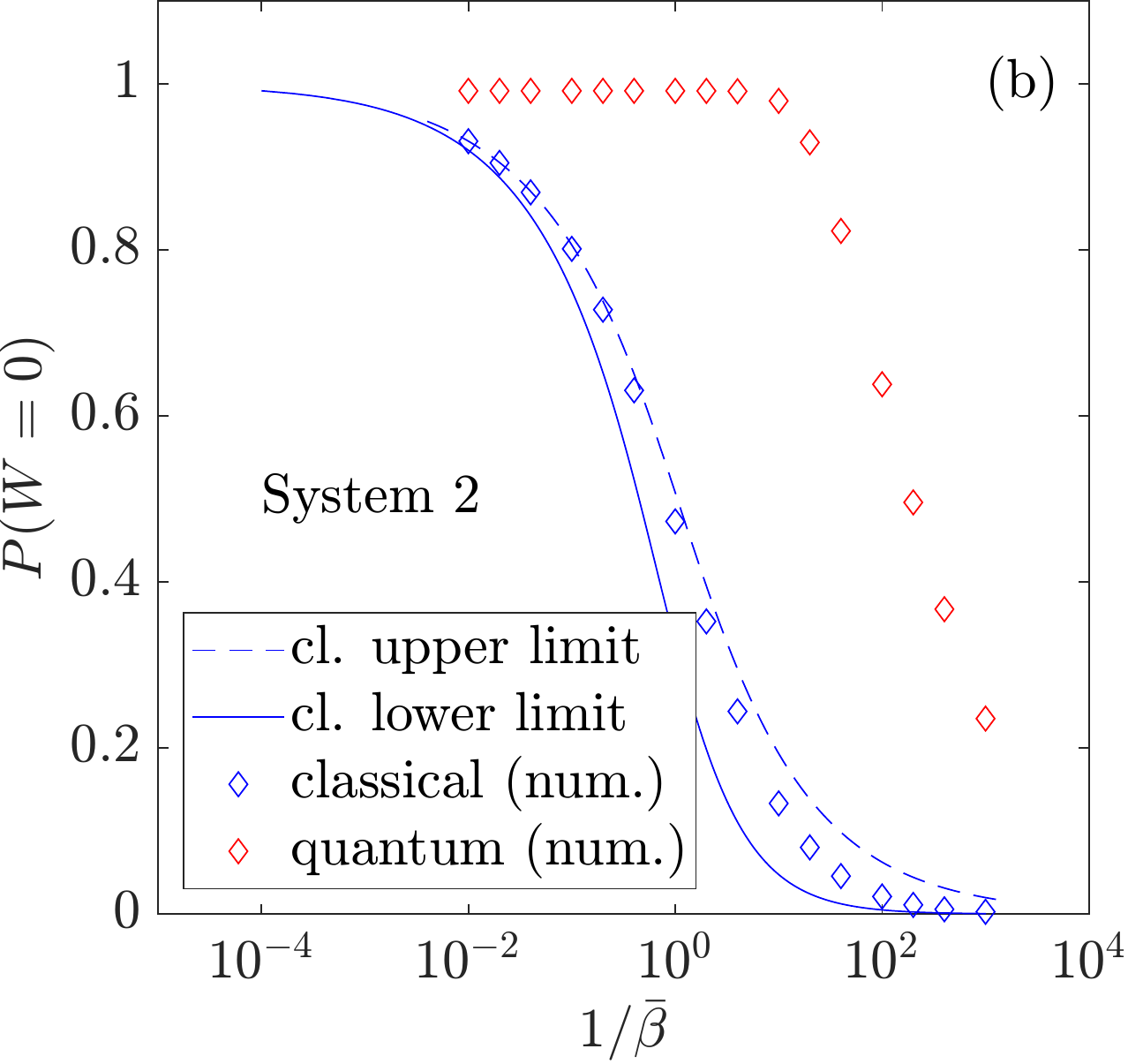}
\caption{Classical (blue) and quantum mechanical (red) probability of no energy change depending on temperature $1/\bar{\beta}$ for (a): \textit{System 1} and for (b): \textit{System 2}. The blue solid lines are given by Eq.~\eqref{eq:P1W0ana} and the dashed line is given by Eq.~\eqref{eq:P2W0ana}. The quantum mechanical calculations were performed for Eq.~\eqref{eq:def}.}
\label{fig:PW0}
\end{figure}

For \textit{System 1} the analytical formula Eq.~\eqref{eq:P1W0ana} matches accurately the numerical simulations of $10^5$ classical particles per point, see Fig.~\ref{fig:PW0}(a). Similarly good results we find for Eqs.~\eqref{eq:P1W0ana} and \eqref{eq:P2W0ana}, which limit the probability $P(W=0)$ for classical particles in \textit{System 2}, see Fig.~\ref{fig:PW0}(b).
As already mentioned in Sec.~\ref{sec:PW} the probability $P(W=0)$ decreases with higher temperatures (small $\bar\beta$) in the classical as well as in the quantum case. But different to the classical case where $P_2^{\rm cl}(W=0) \geq P_1^{\rm cl}(W=0)$ at higher temperatures we find $P_2^{\rm qm}(W=0) < P_1^{\rm qm}(W=0)$, cf. Fig.~\ref{fig:FW}. This is related to the forbidden transitions in \textit{System 1} as discussed at the end of Sec.~\ref{sec:PW}. In addition it seems that in \textit{System 1} the value of $P_1^{\rm qm}(W=0)$ converges at high temperatures to a non-zero plateau. 

\subsection{Probability of no angular momentum change} \label{sec:PL0}
We now calculate the probability of no angular-momentum change. 

\textit{System 1} is radial symmetric at all times, the Hamiltonian is not explicit angular-dependent, so the angular momentum is a conserved quantity and 
\begin{align}
 P_1(\Delta L=0)=1
\end{align} 
is trivial in both classical and quantum mechanics. 

This is different in \textit{System 2} which is only radial symmetric at the beginning and at the end. 
In the classical case all collisions (also those the particle performs with the resting horizontal edges and conserve the energy) change the angular momentum. Except for the occasional one further collisions cannot compensate this change exactly. It follows 
\begin{align}
 P_2^{\rm cl}(\Delta L=0) = 1-e^{-\bar{\beta}}\left[ I_0(\bar{\beta})+I_1(\bar{\beta}) \right] \, , \label{eq:P2L0ana}
\end{align}
with $\bar{\beta}$ defined in Eq.~\eqref{eq:barbeta}. The r.h.s. is identical with that in Eq.~\eqref{eq:P1W0ana} since it represents the probability of no collisions. For details see App.~\ref{sec:AppB}. So $P_2^{\rm cl}(\Delta L=0) \leq P_2^{\rm cl}(W=0)$ is trivial, see Eq.~\eqref{eq:P2W0ana}.

In contrast to this is the quantum mechanical case.
Indeed quantum mechanical transitions which change the parity of angular quantum numbers are forbidden but other transitions $\Delta L = 0,2,4...$ are allowed.
Note, all transitions (also those changing the main quantum number) which conserve the angular momentum quantum number contribute to $P(\Delta L=0)$,
\begin{align}
 P_2^{\rm qm}(\Delta L=0) = \frac{1}{Z^{\rm qm}} \sum_{n,n_0,l_0} e^{-\beta E_{n_0,l_0}} P(n,l_0|n_0,l_0) \, , \label{eq:P2L0qm} 
\end{align}
especially it follows $P_2^{\rm qm}(\Delta L=0) \geq P_2^{\rm qm}(W=0)$, see Eq.~\eqref{eq:PW0qm}.

\begin{figure}
\centering
\includegraphics[width=.48\linewidth]{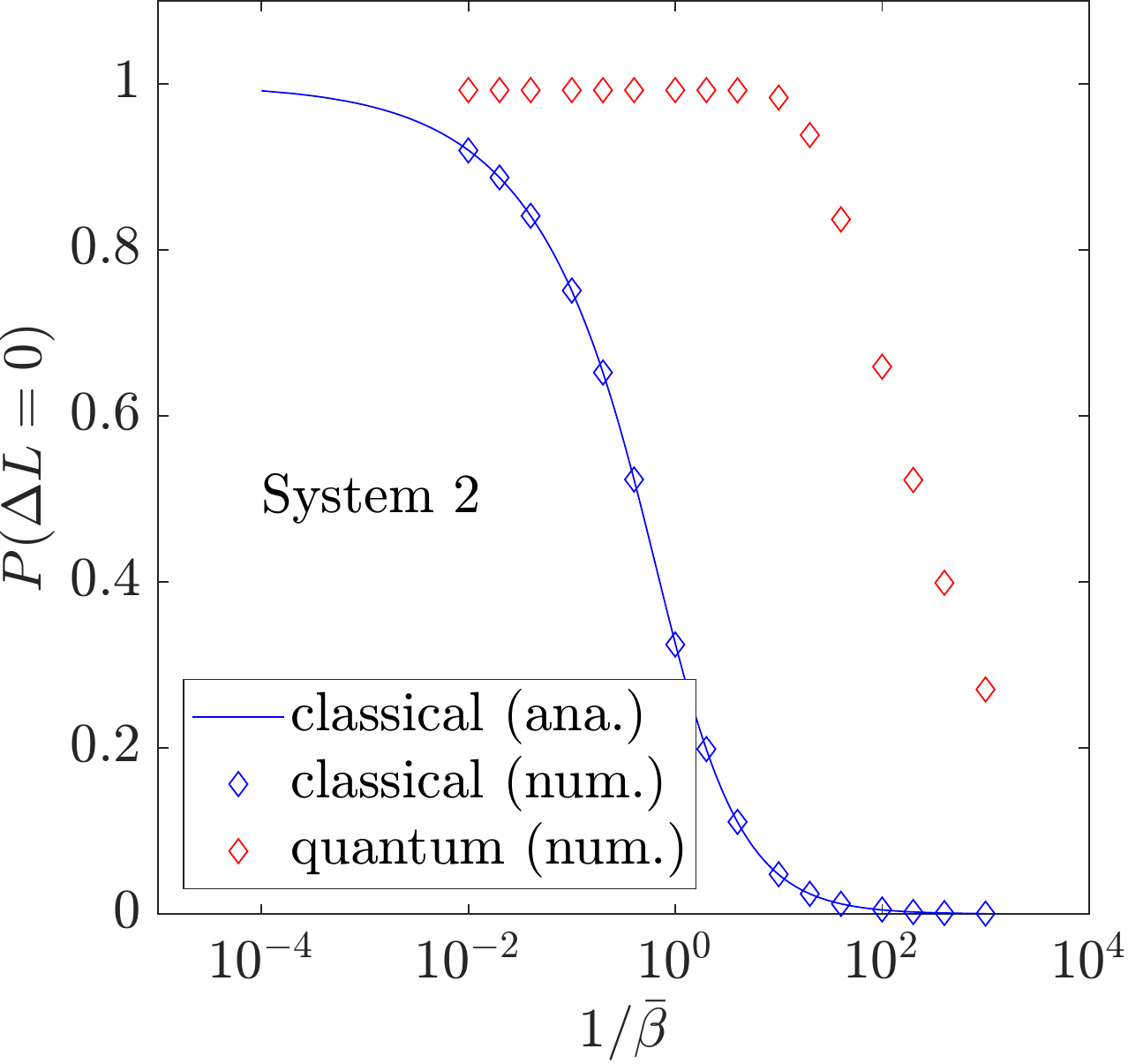} 
\caption{Classical (blue) and quantum mechanical (red) probability of no angular momentum change depending on temperature $1/\bar{\beta}$ for \textit{System 2}. The blue solid line is given by Eq.~\eqref{eq:P2L0ana}. The quantum mechanical calculations were performed for Eq.~\eqref{eq:def}.}
\label{fig:PL0}
\end{figure}

These effects are illustrated in Fig.~\ref{fig:PL0}. Similar as in the previous section formula Eq.~\eqref{eq:P2L0ana} matches accurately the numerical simulations of $10^5$ classical particles per point. Comparing with Fig.~\ref{fig:PW0}(b) in which the numerical results lie above the full line we see our numerical results confirm the expectation $P_2^{\rm cl}(\Delta L=0) \leq P_2^{\rm cl}(W=0)$. Of course, the quantum mechanical numerical results confirm $P_2^{\rm qm}(\Delta L=0) \geq P_2^{\rm qm}(W=0)$. This is trivial since Eqs.~\eqref{eq:PW0qm} and \eqref{eq:P2L0qm} are used for these calculations, respectively. As in Fig.~\ref{fig:PW0}(b) the quantum values for $P_2^{\rm qm}(\Delta L=0)$ decrease for higher temperatures.

%
%


%


\section{Conclusion} \label{sec:conc}

The aim of this paper is to contribute on the field of quantum work statistics. We have picked out two 2--dim billiard systems one of them is a classically integrable breathing circle (\textit{System 1}) and the other one forms a classical-chaotic stadium (\textit{System 2}). 
Classical calculations of trajectories have been performed iteratively. For the quantum mechanical calculations we have had to solve the time-dependent Schrödinger equation. Whereas there is an analytical solution for \textit{System 1} the evolution of wave functions in \textit{System 2} has been solved by the spectral method. 
Using these ingredients the classical conditional probability density $p(E,L|E_0,L_0)$ as well as the quantum mechanical transition probability $P(n,l|n_0,l_0)$ follow which build the basis for statistical analysis. 

So it has been possible to calculate the work distribution for a particle in such systems. Especially the results in the quantum case are of particular interest since already a suitable definition of mechanical work in small quantum systems is controversial.
We find that for higher temperatures the classical and quantum mechanical work distributions converge to each other. Nevertheless, on the one hand the convergence in \textit{System 1} is limited. There are temperature-independent barriers because of angular momentum conservation, e.g. values of work between $0$ and $E_{2,0}-E_{1,0}$ are forbidden. For \textit{System 2} there cannot exist such barriers.  On the other hand a rapidly increasing number of eigenstates has to be calculated. At this point semi-classical methods become of relevance which may build a connection between very fast classical simulations and very cumbersome quantum mechanical calculations. The semi-classical analysis of these systems is part of further research.


Furthermore, we present the results for the probability of no energy or no angular momentum change. Using connections to an exact solvable system analytical formulas are given for these classical probabilities in both systems. These formulas may be applied to other classical billiard systems. For \textit{System 2} it is trivial to see that in the classical case all collisions yield an angular momentum change. Collisions with the horizontal walls did not change the energy. So the probability of no angular momentum change is lower than the probability of no energy change, $P_2^{\rm cl}(\Delta L=0) \leq P_2^{\rm cl}(W=0)$. This is in contrast to the quantum case, $P_2^{\rm qm}(\Delta L=0) \geq P_2^{\rm qm}(W=0)$. Also here semi-classical methods may clarify these fundamental difference.

\begin{acknowledgments}
We would like to thank Andreas Engel, Axel Prüser and the members of the DFG Research Unit FOR2692 for fruitful discussions.
This work has been funded by the Deutsche Forschungsgemeinschaft (DFG, German Research Foundation) -- 397082825. 
\end{acknowledgments}


\appendix

\section{Classical descriptions of System 1 and 2} \label{sec:AppA}

\subsubsection{Radial breathing Circle (\textit{System 1})}

If the particle velocity is larger than the expansion velocity of the circle $|{\bf v}|>u$, the first collision appears at
\begin{align}
 t_1 &= \frac{u R_0 - {\bf r}_0 {\bf v}_0}{v_0^2-u^2} + \Delta(u,R_0,{\bf r}_0,{\bf v}_0) \, , \\
 \Delta(u,R_0,{\bf r}_0,{\bf v}_0) &= \sqrt{\left( \frac{u R_0 - {\bf r}_0 {\bf v}_0}{v_0^2-u^2} \right)^2 + \frac{R_0^2-r_0^2}{v_0^2-u^2}} \, ,
\end{align}
on position 
\begin{align}
 {\bf R}_1 &= {\bf r}_0 + {\bf v}_0 t_1 \, ,
\end{align}
if $t_1<T/2$. If $t_1\geq T/2$ or $|{\bf v}|\leq u$, there is no collision in the expanding phase and the particle position and velocity at $t=T/2$ are
\begin{align}
 {\bf \bar r}_0 &= {\bf r}_0 + {\bf v}_0 \frac{T}{2} \, , \\
 {\bf \bar v}_0 &= {\bf v}_0 \, .
\end{align} 
At each collision during the expanding phase only the radial part of the particle velocity is decreased by $2u$ whereas the angular part stays constant; in the $j$--th collision the velocity ${\bf v}_{j-1}$ changes by 
\begin{align}
 {\bf v}_j &= {\bf v}_{j-1} - 2 \frac{uR_j - {\bf R}_j {\bf v}_{j-1} }{R_j^2} {\bf R}_j \, .
\end{align}
If $v_j^2>u^2$ the next collision time is determined by
\begin{align}
 t_{j+1} &= 2 \frac{u R_j-{\bf R}_j{\bf v}_j}{v_j^2-u^2} \quad j \geq 1 \, .
\end{align}
The collision position is
\begin{align}
 {\bf R}_{j+1} &= {\bf R}_j + {\bf v}_j t_{j+1} \, .
\end{align}
The smallest $j=J$ for which either $v_J^2 \leq u^2$ or $\sum_{n=1}^{J+1} t_{n} > T/2$ is the number of collisions in the expanding phase; so for the particle position and velocity at $t=T/2$ follow
\begin{align}
 {\bf \bar r}_0 &= {\bf R}_J + {\bf v}_J \left(\frac{T}{2} - \sum_{n=1}^{J} t_{n} \right) \, , \\
 {\bf \bar v}_0 &= {\bf v}_J \, .
\end{align} 
In the contracting phase ($u \rightarrow \bar u = -u$, $\bar R_0 = R_0 + u \frac{T}{2}$) the first collision appears at
\begin{align}
 \bar t_1 &= \begin{cases} \frac{\bar u \bar R_0 - {\bf \bar r}_0{\bf \bar v}_0}{\bar v_0^2-\bar u^2} + \Delta(\bar u,\bar R_0,{\bf \bar r}_0,{\bf \bar v}_0) & \bar v_0^2> \bar u^2 \\ \frac{\bar u \bar R_0 - {\bf \bar r}_0{\bf \bar v}_{0}}{\bar v_0^2-\bar u^2} - \Delta(\bar u,\bar R_0,{\bf \bar r}_0,{\bf \bar v}_0) & \bar v_0^2<\bar u^2 \\ \frac{\bar R_0^2-\bar r_0^2}{2\left({\bf \bar r}_0 {\bf \bar v}_0-\bar R_0 \bar u \right)} & \bar v_0^2=\bar u^2 \, , \end{cases}
\end{align}
on position 
\begin{align}
 {\bf \bar R}_1 &= {\bf \bar r}_0 + {\bf \bar v}_0 \bar t_1 \, ,
\end{align}
if $\bar t_1 \leq \frac{T}{2}$. In the case $\bar t_1 > \frac{T}{2}$ there is no collision in the contracting phase and the final particle position and velocity are
\begin{align}
 {\bf r}_f &= {\bf \bar r}_0 + {\bf \bar v}_0 \frac{T}{2} \, , \\
 {\bf v}_f &= {\bf \bar v}_0 \, .
\end{align} 
At each collision during the contraction phase only the radial part of the particle velocity is increased by $2u$ whereas the angular part stays constant; in the $k$--th collision the velocity ${\bf \bar v}_{k-1}$ changes by 
\begin{align}
 {\bf \bar v}_k &= {\bf \bar v}_{k-1} - 2 \frac{\bar u \bar R_k - {\bf \bar R}_k {\bf \bar v}_{k-1} }{\bar R_k^2} {\bf \bar R}_k \, .
\end{align}
The next collision time is determined by
\begin{align}
 \bar t_{k+1} &= 2 \frac{\bar u \bar R_k-{\bf \bar R}_k{\bf \bar v}_k}{\bar v_k^2- \bar u^2} \quad k \geq 1 \, ,
\end{align}
on position
\begin{align}
 {\bf \bar R}_{k+1} &= {\bf \bar R}_k + {\bf \bar v}_k \bar t_{k+1} \, .
\end{align}
The lowest $k=K$ for which $\sum_{n=1}^{K+1} \bar t_{n} > T/2$ is the number of collisions in the contracting phase; so the final particle position and velocity are
\begin{align}
 {\bf r}_f &= {\bf \bar R}_K + {\bf \bar v}_K \left(\frac{T}{2} - \sum_{n=1}^{K} \bar t_{n} \right) \, , \\
 {\bf v}_f &= {\bf \bar v}_K \, .
\end{align} 

\subsubsection{Horizontal breathing Stadium (\textit{System 2})}

In the following we give the explicit formulas for the first collision time and position in the case $x_0 \geq 0$ and $y_0 \geq 0$ assuming the first collision takes place in the expanding phase. For all other cases as well as for further collisions and collisions in the contraction phase the explicit formulas can be derived in a similar way.

The initial velocity is ${\bf v}_0=\left(\begin{smallmatrix} a_0 \\ b_0 \end{smallmatrix}\right)$. If $b_0=0$ and $|a_0|<u$ no collision will happen in the expansion phase. For $b_0=0$ and $a_0<-|u|$ the first collision appears on the left half circle, for $a_0>|u|$ (independent of $b_0$) the first collision appears on the right half circle. In all other cases the location of the first collision (left or right half circle or static top or bottom line) depends on three characteristic times:
\begin{enumerate}
 \item the escape time from the right half circle $t^{\rm (I)}= \frac{x_0}{|u|-a_0}$, only relevant if $a_0<|u|$,
 \item the collision time on the top ($b_0>0$) or bottom ($b_0<0$) line $t^{\rm (II)}=\frac{\sign(b_0) R_0-y_0}{b_0}$, and
 \item the entry time to the left half circle $t^{\rm (III)} =\frac{-x_0}{|u|+a_0}>t^{\rm (I)}$, only relevant if $a_0<-|u|$.
\end{enumerate}
If $t^{\rm (II)}<t^{\rm (I)}$ or $a_0>u$ the first collision will appear on the right half circle at time $t_1$ on position ${\bf R}_1$ and the velocity changes to ${\bf v}_1$:
\begin{widetext}
\begin{align}
 t_1 &= \frac{-(a_0-u)x_0-b_0y_0}{(a_0-u)^2+b_0^2}+\sqrt{\left[\frac{-(a_0-u)x_0-b_0y_0}{(a_0-u)^2+b_0^2}\right]^2+\frac{R^2-x_0^2-y_0^2}{(a_0-u)^2+b_0^2}} \\
 {\bf R}_1 &= \begin{pmatrix} X_1 \\ Y_1 \end{pmatrix} = {\bf r}_0 + {\bf v}_0t_1 \\
 {\bf v}_1 &= \frac{1}{R_0^2} \begin{pmatrix} uR_0^2 + (a_0-u)[Y_1^2-(X_1-ut_1)^2]-2b_0(X_1-ut_1)Y_1 \\ b_0[(X_1-ut_1)^2-Y_1^2]-2(a_0-u)(X_1-ut_1)Y_1 \end{pmatrix} \label{eq:Appv1}  
\end{align}
\end{widetext}
Else if $t^{\rm (I)}<t^{\rm (II)}<t^{\rm (III)}$ the first collision will appear on the static top or bottom line at time 
\begin{align}
 t_1 &= t^{\rm (II)}
\end{align}
on position
\begin{align}
 {\bf R}_1 &= {\bf r}_0 + {\bf v}_0 t_1
\end{align}
and the velocity changes to
\begin{align}
 {\bf v}_1 &= \begin{pmatrix} a_0 \\ -b_0 \end{pmatrix} \, . \label{eq:Appv2}
\end{align}
Else the first collision will appear on the left half circle at time $t_1$ on position ${\bf R}_1$ and the velocity changes to ${\bf v}_1$:
\begin{widetext}
\begin{align}
 t_1 &= t^{\rm (III)} - \frac{(y_0+b_0t^{\rm (III)})b_0}{(a_0+u)^2+b_0^2}+\sqrt{\left[\frac{(y_0+b_0t^{\rm (III)})b_0}{(a_0+u)^2+b_0^2}\right]^2+\frac{R^2-(y_0+b_0t^{\rm (III)})^2}{(a_0+u)^2+b_0^2}} \\
 {\bf R}_1 &= \begin{pmatrix} X_1 \\ Y_1 \end{pmatrix} = {\bf r}_0 + {\bf v}_0t_1 \\
 {\bf v}_1 &= \frac{1}{R_0^2} \begin{pmatrix} - uR_0^2 + (a_0+u)[Y_1^2-(X_1+ut_1)^2]-2b_0(X_1+ut_1)Y_1\\ b_0[(X_1+ut_1)^2-Y_1^2]-2(a_0+u)(X_1+ut_1)Y_1 \end{pmatrix} \, . \label{eq:Appv3}
\end{align}
\end{widetext}

After each collision the new position and the new velocity determine on which wall (left/right half circle or top/bottom line) the next collision will happen. The velocity change depends on that wall, see Eqs.~\eqref{eq:Appv1}, \eqref{eq:Appv2} and \eqref{eq:Appv3} respectively. A simple closed form of $t_1$, ${\bf R}_1$ and ${\bf v}_1$ does not exist.

Nevertheless, similar considerations as for the first collision may be done for further collisions up to $t=T/2$ as well as for collisions in the contracting phase. 

\section{Classical probability of no energy change and no angular momentum change} \label{sec:AppB}

Now, we introduce a static circular billiard system with radius $R_0$ in thermal equilibrium with a heat bath at inverse temperature $\beta$. At $t=0$ system and bath are decoupled and we remove the walls. This System is referred to as \textit{System 0}. The detailed relations with \textit{System 1} and \textit{2} are explained below.

In \textit{System 0}, all particle perform a rectilinear motion. The distance to the center of the initial circle is given by 
\begin{align}
 r_T(r_0,\varphi;vT) &= \sqrt{(r_0+vT \cos\varphi)^2+(vT)^2\sin^2\varphi} 
\end{align}
depending on the initial distance $r_0$ (linearly distributed), the launching angle $\varphi$ (uniformly distributed) and the length of path $vT$.
So $r_T$ is a stochastic variable, too
\begin{align}
 p_0(r_T;vT) &= \left\langle \delta\left( r_T(r_0,\varphi;vT)-r_T \right) \right\rangle_{r_0,\varphi} \\
 \left\langle ( \cdot ) \right\rangle_{r_0,\varphi}&= \int\limits_{0}^{R_0} dr_0 \int\limits_{0}^{2\pi} d\varphi \, ( \cdot ) p(r_0) p(\varphi) \, .
\end{align}
This double-integral can be performed analytically
\begin{widetext}
\begin{align}
 \label{eq:Prt}p_0(r_T;vT) &= \begin{cases} \frac{2r_T}{R_0^2} & 0 \leq r_T \leq {\rm max}\{0;R_0-vT\} \\ \frac{r_T}{R_0^2} - \frac{2r_T}{\pi R_0^2} \arctan\left\{ \frac{-R_0^2+r_T^2+(vT)^2}{\sqrt{-r_T^4-(R_0-vT)^2R_0^2+2r_T^2\left[R_0^2+(vT)^2\right]}} \right\} & {\rm max}\{0;R_0-vT\} \leq r_T<R_0+vT \\ 0 & \text{otherwise.} \end{cases} 
\end{align}
\end{widetext}
In a canonical ensemble at the beginning the velocity of a particle $v(E)$ is Maxwell distributed. The probability of presence of this particle inside the initial circle (with radius $R_0$) is given by
\begin{align}
 P_0(r_T \leq R_0) &= \int\limits_{0}^{\infty} dE \, \beta e^{-\beta E} \int\limits_{0}^{R_0} dr_T \, p_0(r_T;v(E)T) \nonumber \\
 &= 1-e^{-\bar{\beta}}\left[ I_0(\bar{\beta})+I_1(\bar{\beta}) \right] \, , \\
 \bar{\beta}&= \frac{m R_0^2}{T^2}\beta  \nonumber \, ,
\end{align}
where $I_0$ and $I_1$ are modified Bessel's functions of the first kind.

The probability $P_0(r_T \leq R_0)$ is equal to the probability in \textit{System 1} and \textit{2} that no collisions were performed, so
\begin{align}
 P_0(r_T \leq R_0) &= P_1^{\mathrm{cl}}(W=0) \, , \\
 P_0(r_T \leq R_0) &= P_2^{\mathrm{cl}}(\Delta L=0) \leq P_2^{\mathrm{cl}}(W=0) \, .
\end{align}
Especially for \textit{System 1} any collision or combinations of collisions change the energy of the classical velocity except for the occasional one. Only fine tuned combinations of starting positions and launching angles are able to compensate energy loss and energy gain exactly. 

Now, we consider the probability density of \textit{System 0} again, Eq.~\eqref{eq:Prt}. Imagine a lot of circles upward and downward the initial circle with midpoints on the $y$-axis. All these circles have the same radius $R_0$ and are tangent to their (both) neighbours. With a canonical velocity distribution the probability of presence in the $k$-th upper or lower circle can be simplified to

\begin{align}
 \label{eq:B8}P_{00} &= P_0(r_T \leq R_0) \\
 P_{0k} &= \bar{\beta} \int\limits_{2k-2}^{2k} dx \, x e^{-\frac{\bar{\beta}}{2}x^2} \nonumber \\ & \quad \times \int\limits_{(2k-1)R_0}^{(1+x)R_0} dr_T \, p_0(r_T;x R_0) \cdot \Pi_k(r_T) \, ,
\end{align}
where $\Pi_k(r_T)$ is the weight that states which particle fraction is inside the $k$--th upper or lower circle
\begin{align}
 &\Pi_k(r_T)= \nonumber \\ & \frac{2}{\pi} \arcsin\left( \frac{\sqrt{16k^2R_0^4-(r_T^2-R_0^2-4k^2R_0^2)^2}}{4kr_TR_0} \right) \, . \label{eq:Pigeo}
\end{align}

A reflection in \textit{System 2} with the horizontal resting walls can be projected as a mirrored trajectory which entered an upward or a downward stadium.
Therefore we consider a particle in \textit{System 2} which collides only on the horizontal walls. In \textit{System 0} the same particle has to be located in the initial circle or in one of the imagined circles (upward/downward). Note, the inversion does not apply. So its probability of presence in one of these circles is an upper limit for $P_2^{\mathrm{cl}}(W=0)$ in \textit{System 2}. So
\begin{align}
 P_2^{\mathrm{cl}}(W=0) \leq \sum_{k=0}^{\infty} P_{0k} \, . \label{eq:P2W0as}
\end{align}

Substituting $y=r_T/R_0$, Eqs.~\eqref{eq:B8}-\eqref{eq:P2W0as} combines to the upper limit Eq.~\eqref{eq:P2W0ana}.

%

\bibliography{references}

\end{document}